\documentclass[twocolumn, usenatbib]{mnras}

\usepackage{amsmath}
\usepackage{float}

\usepackage[usenames, dvipsnames]{color}
\usepackage{graphicx}
\usepackage{verbatim}

\usepackage{hyperref}
\hypersetup{
    pdfnewwindow=true,      
    colorlinks=true,       
    linkcolor=blue,          
    citecolor=cyan,        
    filecolor=blue,      
    urlcolor=blue           
}

\def\prd{PRD}
\def\apj{ApJ}                 
\def\apjl{ApJL}               
\def\apjs{ApJS}               
\def\mnras{MNRAS}             
\def\aap{A\&A}                
\def\nat{Nature}              
\def\araa{ARA\&A}             

\newcommand\lsim{\mathrel{\rlap{\lower4pt\hbox{\hskip1pt$\sim$}}
        \raise1pt\hbox{$<$}}}
\newcommand\gsim{\mathrel{\rlap{\lower4pt\hbox{\hskip1pt$\sim$}}
        \raise1pt\hbox{$>$}}}

\def\eff{\rm{eff}}
\def\tot{\rm{bol}}

\def\Rin{R_{\rm{d}}}
\def\em{\rm{em}}

\def\Msun{ M_{ \rm{\odot} } }

\begin{document}
\label{firstpage} 
\pagerange{\pageref{firstpage}--\pageref{lastpage}}

\title[A Lighthouse in the Dust]{Lighthouse in the Dust: Infrared Echoes of Periodic Emission from Massive Black Hole Binaries\thanks{We dedicate this work to the memory of Arlin Crotts, an expert on supernova light echoes who passed away on November 19, 2015.}}

\author[D. J. D'Orazio, Z. Haiman]{Daniel J. D'Orazio$^{1}$\thanks{Einstein Fellow; daniel.dorazio@cfa.harvard.edu}, Zolt\'an~Haiman$^{2,3}$\\
     $^{1}$Astronomy Department, Harvard University, 60 Garden Street, Cambridge, MA 02138, USA \\
     $^{2}$Department of Astronomy, Columbia University, 550 West 120th Street, New York, NY 10027, USA\\     $^{3}$Department of Physics, New York University, New York, NY 10003, USA\\
     }

\maketitle

\begin{abstract}
The optical and UV emission from sub-parsec massive black hole binaries
(MBHBs) in active galactic nuclei (AGN) is believed to vary periodically, on
timescales comparable to the binary’s orbital time. If driven by accretion
rate fluctuations, the variability could be isotropic. If dominated by
relativistic Doppler modulation, the variability should instead be
anisotropic, resembling a rotating forward-beamed lighthouse. We consider the
infrared (IR) reverberation of either type of periodic emission by pc-scale
circumbinary dust tori. We predict the phase and amplitude of IR variability
as a function of the ratio of dust light crossing time to the source
variability period, and of the torus inclination and opening angle. We
enumerate several differences between the isotropic and anisotropic cases.
Interestingly, for a nearly face-on binary with an inclined dust torus, the
Doppler boost can produce IR variability without any observable optical/UV
variability. Such orphan-IR variability would have been missed in optical
searches for periodic AGN. We apply our models to time-domain WISE IR data
from the MBHB candidate PG 1302-102 and find consistency with dust
reverberation by both isotropically emitting and Doppler-boosted sources in
the shorter wavelength W1-W2 ($ 2.8 \rightarrow 5.3 \mu$m) bands. We constrain
the dust torus to be thin (aspect ratio $\sim 0.1$), with an inner radius at
1-5 pc. More generally, our dust echo models will aid in identifying new MBHB
candidates, determining their nature, and constraining the physical properties
of MBHBs and their dust tori.
\end{abstract}

\begin{keywords} 
quasars: supermassive black holes --  quasars: individual: PG1302-102
\end{keywords}

\section{Introduction} 

Massive black holes (MBHs) exist at the centers of most, if not all, galaxies
\citep{kr95, KormendyHo2013}. Galactic mergers can deliver MBHs, as well as an
ample supply of gas \citep{BH1992, Barnes:1996, Barnes:2002,
Mayer:2013:MBHBGasRev}, to the centers of newly coalesced galaxies where the
MBHs form a binary. The interaction of massive black hole binaries (MBHBs)
with gas and surrounding stars can drive the pair to sub-pc separations where
gravitational radiation reaction drives the binary to coalescence
\citep{Begel:Blan:Rees:1980}. Characterization of the population of such sub-pc 
binaries, through present electromagnetic (EM), and future gravitational
wave (GW) channels will provide a powerful tool for understanding the mutual
build-up of galaxies and their central black holes
\citep[\textit{e.g.}][]{KormendyHo2013}, the dynamics of gas and stars in
galactic nuclei \citep[\textit{e.g.}][]{MerrittMilos:2005:LRR}, and the 
low-frequency gravitational wave sky \citep[\textit{e.g.}][]{KocsisSesana:2011,
Shannon:2015, Arzoumanian:2015:SGWB, Ravi+2015}.

Electromagnetic signatures of MBHBs can arise from their interaction
with gas.  Hydrodynamical simulations of gas discs surrounding compact
MBHBs show that accretion rates onto a binary are generally comparable
to the accretion rates onto a single black hole of an equivalent mass
\citep{ShiKrolik:2012, DHM:2013:MNRAS, Farris:2014, ShiKrolik:2015,
  MunozLai:2016}. Binary accretion can also be identified by its
unique signatures. Depending on the ratio of BH masses, $q \equiv
M_2/M_1 \leq 1$, the accretion induced emission can be
periodically modulated at predictable frequencies
(\citealt{Hayasaki:2007},
\citealt{MacFadyen:2008},
\citealt{Cuadra:2009},
\citealt{Roedig:2012:Trqs},
\citealt{Noble+2012},
\citealt{ShiKrolik:2012},
\citealt{DHM:2013:MNRAS},
\citealt{Farris:2014},
\citealt{Gold:GRMHD_CBD:2014},
\citealt{Gold:GRMHD_CBDII:2014},
\citealt{Dunhill+2015},
\citealt{ShiKrolik:2015},
\citealt{Farris:2015:Cool},
\citealt{D'Orazio:CBDTrans:2016},
\citealt{MunozLai:2016}).

Tell-tale luminosity variations caused by an accreting MBHB can also occur due
to special relativity alone. For a binary orbiting at relativistic speeds,
relativistic Doppler boosting causes luminosity variations at the binary
orbital period \citep{PG1302MNRAS:2015a,   PG1302Nature:2015b}.  For binaries
with disparate mass ratios, $q \lsim 0.05$, accretion has been found to be
steady and dominated by the smaller, secondary BH \citep{DHM:2013:MNRAS,
Farris:2014}. In this case, for a sufficiently compact binary, Doppler
boosting of the faster-moving secondary BH's emission is expected to be the
dominant source of variability. Even near-equal mass binaries, for which high-
amplitude accretion variability is expected, may emit steadily in their rest
frame.  This would be the case if the minidiscs surrounding each BH act as
buffers between accretion rate variations at the edge of the minidisc and
regions further inside which dominate the emission \citep{TanakaHaiman2013,
Roedig+Krolik+Miller2014}. In this case too, Doppler boosting may be the
dominant source of variability from accreting, close MBHBs.\footnote{For
equal-mass binaries the two BHs would be seen out of phase, canceling the net
Doppler-boost, unless one BH outshines the other,  \textit{e.g.}, for
eccentric binaries \citep{MunozLai:2016}.}

Both accretion rate variability and the Doppler boost scenario for periodic
emission from MBHBs have recently been developed to interpret the MBHB
candidate PG~1302-102 \citep{Graham+2015a, PG1302MNRAS:2015a, PG1302-Maria,
PG1302Nature:2015b, KunPG1302:2015}, a bright $z=0.3$ quasar exhibiting
nearly sinusoidal periodicity in the V-band continuum. Given the measured
binary mass, period, and spectral slope of PG~1302-102,
\cite{PG1302Nature:2015b} showed that the observed amplitude of variability in
the V-band and at ultra-violet (UV) wavelengths is consistent with that
expected for Doppler boosting of emission from an accretion disc around the
secondary BH. Further confirmation of the Doppler-boosting model for
PG~1302-102 requires continued, long-term monitoring of the system at optical
and UV wavelengths.

Measurements at other wavelengths can provide additional clues to the nature
of the central engine of PG~1302-102, as well as of other sub-pc MBHB
candidates \citep{Graham+2015b, Charisi+2016,   Liu:7pc:2015, Zheng:2015,
LiWang:2016}.  Here we focus on time-domain observations of PG~1302-102 in the
infrared (IR). \citet[][hereafter J15]{Jun:2015} have recently analyzed data
from the Wide-field Infrared Survey Explorer (WISE) satellite to report a
periodicity in the IR continuum of PG~1302-102 which is consistent with the
optical period, but with a phase lag of $335 \pm 153$ days in the W1 band (3.4
$\mu$m) and $524 \pm 148$ days in the W2 band (4.6 $\mu$m).  J15 attribute
this phase lag to reverberation of the optical/UV continuum of PG~1302-102 by
a surrounding dusty torus at $\sim$pc distances from the illuminating source.
Another interesting feature -- not discussed by J15 -- is the diminished
amplitude ($\sim8\%$) of the IR fluctuations relative to the optical/UV
variability ($\sim13\%$).

In this work, we develop a toy model to interpret the findings of J15 and, in
general, reverberated IR emission from a periodically variable central
optical/UV source.  While dust echoes of variable AGN have been considered in
the past \citep[\textit{e.g.}][]{Barvainis:1992, HoenigKishimoto:2011,
Koshida+2014, Honig+2017}, the expected periodic nature of the brightness
fluctuations of MBHBs, on tractable timescales of years, makes these systems
unique.  We here examine reverberation of an isotropically ``pulsating''
source, mimicking expectations from fluctuating accretion onto a MBHB.  We
contrast this scenario with reverberation of an anisotropically varying
source, as expected from Doppler-boosted emission from a binary along its
orbit.  Such a source illuminates the surrounding dust similarly to a 
forward-beamed lighthouse, sweeping around at the binary's orbital frequency. 
To our knowledge, reverberation from such a ``rotating light-house'' has not been
previously explored.  While in this paper we focus on dust echoes in the
context of binary AGN, our models can be applied to other reverberating
systems with a periodic, Doppler-boosted central engine.

We employ simple geometric models of a dusty torus that is optically
thick in the optical/UV bands, but optically thin to its own IR dust
emission, and is centered on the emitting source.  We take into
account the relative light travel time to different parts of the
torus, and identify the ratio of dust light crossing time to the
source variability period as the key parameter, setting the amplitude
and phase of the reverberated periodic IR light curve.

We enumerate several differences between the echoes of anisotropic
sources from their isotropic analogues.  We find, for example, that
the phase lags of the IR light curves in the two cases differ by a
quarter cycle.  Most importantly, for the Doppler boosted emission
models, we find that, depending on the relative inclination angles of
the binary's orbital plane and of the symmetry plane of the dusty
torus, variability can be present in both optical and IR, or in either
one band but not the other. This means that present searches for MBHB
candidates in optical surveys may have missed some candidates, and
motivates extending such a search to include IR bands.

As an example of the utility of our models, we apply them to
interpret the IR emission from the MBHB candidate PG~1302-102. We find that
the shorter wavelength ($\lsim 2.8 \rightarrow 5.3 \mu$m) IR emission from
PG~1302-102, for which long term time series data exists, is consistent with
reverberation of either a Doppler-boosted or an isotropically varying central
source.  

While there are not enough epochs in the longer wavelength W3 ($12
\mu$m) and W4 ($22 \mu$m) WISE bands to fit reverberated light curve models,
we use them to fit an IR spectral energy distribution (SED), allowing an
estimate of the total reverberated IR luminosity, which places additional
constraints on the dust radius and opening angle. Including the W3 band in the
SED fitting, we find that the Doppler case is disfavoured and only nearly
face-on dust tori illuminated by an isotropically varying source are allowed.
We find that the bright emission in the W4 band is inconsistent with dust
reverberation by the central source, and must be produced by another
mechanism, such as cooler dust farther from the nucleus.

In the cases consistent with periodic dust reverberation, the dusty
torus has to be thin; the size of the torus differs depending on the SED fit
but has to be between 1-5 pc.

If further monitoring of longer W3 and W4 band
emission from PG 13012-102 can confirm that W3 emission is dominated by
reverberation by the central source, then the Doppler-boosted source may be
ruled out. Otherwise, the cases of an isotropic and Doppler-boosted source
are indistinguishable with current data. We find two main differences between
the predictions of each. First, the Doppler-boosted scenario excludes dust
tori that are within $\sim 55$ degrees of face on, while the isotropic
scenario can allow face-on dust tori. However, the angular size of the
reverberated-IR emitting region in PG~1302-102 is much too small for direct IR
imaging to test this prediction. Second, each scenario predicts different dust
torus radii. Both continued monitoring of the IR and optical/UV light curves
and measurement of the reverberated-IR spectral energy distribution will
narrow the constraints on the dust torus size, allowing differentiation
between the Doppler boost and isotropic models.

The rest of this paper is organized as follows. We proceed in \S\ref{S:Model}
by describing the MBHB and dust system. In \S\ref{S:Derivation} we develop
models for IR emission from a dust region heated by periodically-varying
isotropic and Doppler-boosted MBHB continua. In \S\ref{S:PDs} we present
analytic models to explore the parameter dependencies and their consequences,
differentiating between isotropic and Doppler-boost scenarios. In
\S\ref{S:PG1302} we apply our models to the MBHB candidate PG~1302-102. In
\S\ref{S:Discussion} we summarize our findings and the limitations and
possible extensions of the simple model developed here, and end by offering
our conclusions.

\section{Model Setup}
 \label{S:Model}
 \subsection{Dust Torus}

Parsec-scale dust structures surrounding the central engines of active
galactic nuclei (AGN) are ubiquitous. The unification of type I (unobscured)
and type II (obscured) AGN posits that the difference in AGN types is the
viewing angle relative to an obscuring dust torus \citep{Antonucci:1993,
KrolikBegelman:1988}. This dust, heated by the central AGN source, emits in
the IR and its properties have been investigated by high-resolution IR
imaging as well as by modeling the IR SED.

High resolution IR imaging puts an upper limit of a few parsecs on the size of
the emitting dust region \citep[see, e.g.][and references
therein]{Elitzur:2006}, while a wide variety of models of optically thin and
thick, spatially smooth and clumpy dust distributions of various dust
geometries and compositions have been employed to model the IR SEDs of AGN
\citep[see, e.g. the review by][and references therein]{Netzer:2015:rev}.
Neither imaging nor SED fitting, however, uniquely determines the dust
properties or geometry.

We do not attempt to reproduce the full SEDs of AGN dust tori. Instead, our
goal is to understand the basic properties of IR dust echoes that results from
heating by a \textit{periodically} variable optical/UV continuum, using a
model as simple as possible. To facilitate this goal, we assume that the dust
structure is centered on the illuminating source and absorbs all incident
optical/UV radiation in a thin inner shell. We assume that the dust is
optically thin to its own IR emission and consists of a species with a single
grain size.

\subsection{Central MBHB Source} 

The central source of optical/UV continuum which heats the dust and
causes it to emit in IR is accretion-induced emission from a compact
MBHB. For the expected BH masses ranging from $10^6 \rightarrow
10^{10} \Msun$, the thermal emission from a steady-state accretion
disc \citep{SS73} has a modified blackbody spectrum that peaks in the
X-rays (lower-mass BHs) to the optical (higher-mass BHs). This
radiation can be efficiently absorbed to heat $\mu$m size dust
particles (see \S \ref{S:FISOderivation}).

Here we are interested in the effect of periodically variable emission
from the MBHB. The simplest case is that of an isotropic, sinusoidally
pulsating source,
\begin{equation}
F^{\rm{Iso}}_{\nu} = F^0_{\nu}\left[ 1 + A \sin{\Omega t } \right],
\label{Eq:Fsrc_ISO}
\end{equation}
where $F^0_{\nu}$ is the average specific flux, $A$ is the amplitude and
$P=2\pi / \Omega$ the period of the modulation, and the initial phase is taken
to be zero. We note that, due to limb-darkening and geometric
foreshortening, emission from an accretion disk is likely not spatially
isotropic \citep[\textit{e.g.}][and references therein]{NamekataUmemura:2016},
however, we employ the isotropic case as a toy model for binary accretion-
induced variability, and more importantly, this choice serves as a control
with which to compare to the Doppler-boosted case, which exhibits a distinct,
time-dependent emission anisotropy.

Relativistic Doppler boost modulates the observed flux as
\begin{equation}
F^{\rm{Dop}}_{\nu} = \frac{F^0_{\nu}}{\left[\gamma\left( 1 - \frac{v^{\rm{obs}}_{||}}{c}\right)\right]^{3-\alpha_{\nu}}},
\label{Eq:Dop1}
\end{equation}
where we assume that the emission is intrinsically steady and emanates from a
single BH (the smaller, faster BH in the pair; this assumption is justified
for PG~1302-102 in \citealt{PG1302Nature:2015b}), $c$ is the speed of light,
$v^{\rm{obs}}_{||}$ is the projection of the source velocity along the
observer's line of sight, $\gamma = \left[ 1   - (v_s/c)^2 \right]^{-1/2}$ is
the Lorentz factor of the secondary which orbits at speed $v_s =
(1+q)^{-1}\sqrt{GM/a}$ for binary with mass $M$ and separation
$a$,\footnote{Not to be confused with dust grain radius $a_{\rm{eff}}$
introduced below.} and we assume throughout, for simplicity, that the binary
is on a circular orbit. The variability period is the binary orbital period
\begin{equation} 
P = \frac{2 \pi a^{3/2}}{\sqrt{GM}} \label{Eq:BPrd}.
\end{equation}
Eq. (\ref{Eq:Dop1}) further assumes that the rest-frame specific flux is a power-law
$F^0_{\nu} \propto \nu^{\alpha_{\nu}}$ with slope $\alpha_{\nu} \equiv
d\rm{ln}F_{\nu} / d \rm{ln}\nu$ at the observed frequency. 

If the source is observed over all frequencies, as is approximately the case
for dust absorbing a spectrum of optical/UV emission, then we emphasize here
that the form of the source spectrum is not relevant. Using the Lorentz
invariant quantity $F_{\nu}/\nu^3$ to integrate over all frequencies,
the total observed flux becomes
\begin{equation}
F^{\rm{Dop}} = \frac{F^0}{\left[\gamma\left( 1 - \frac{v^{\rm{obs}}_{||}}{c}\right)\right]^{3-\alpha_{\tot}} },
\label{Eq:DopAll}
\end{equation}
where, as long as we integrate over the entire source spectrum, we must have
$\alpha_{\tot}=-1$. However, we introduce the $\alpha_{\tot}$ notation
because we will also use $\alpha_{\tot}=4$ for the purposes of analytic
approximation below.

Throughout we use the $\alpha_{\nu}$ notation when considering 
``in-band'' observations of Doppler-modulated optical/UV emission, and the
$\alpha_{\tot}$ notation when considering the spectrum-integrated emission
seen by the dust.

\begin{figure}
\begin{center}$
\begin{array}{c}
\includegraphics[scale=0.4]{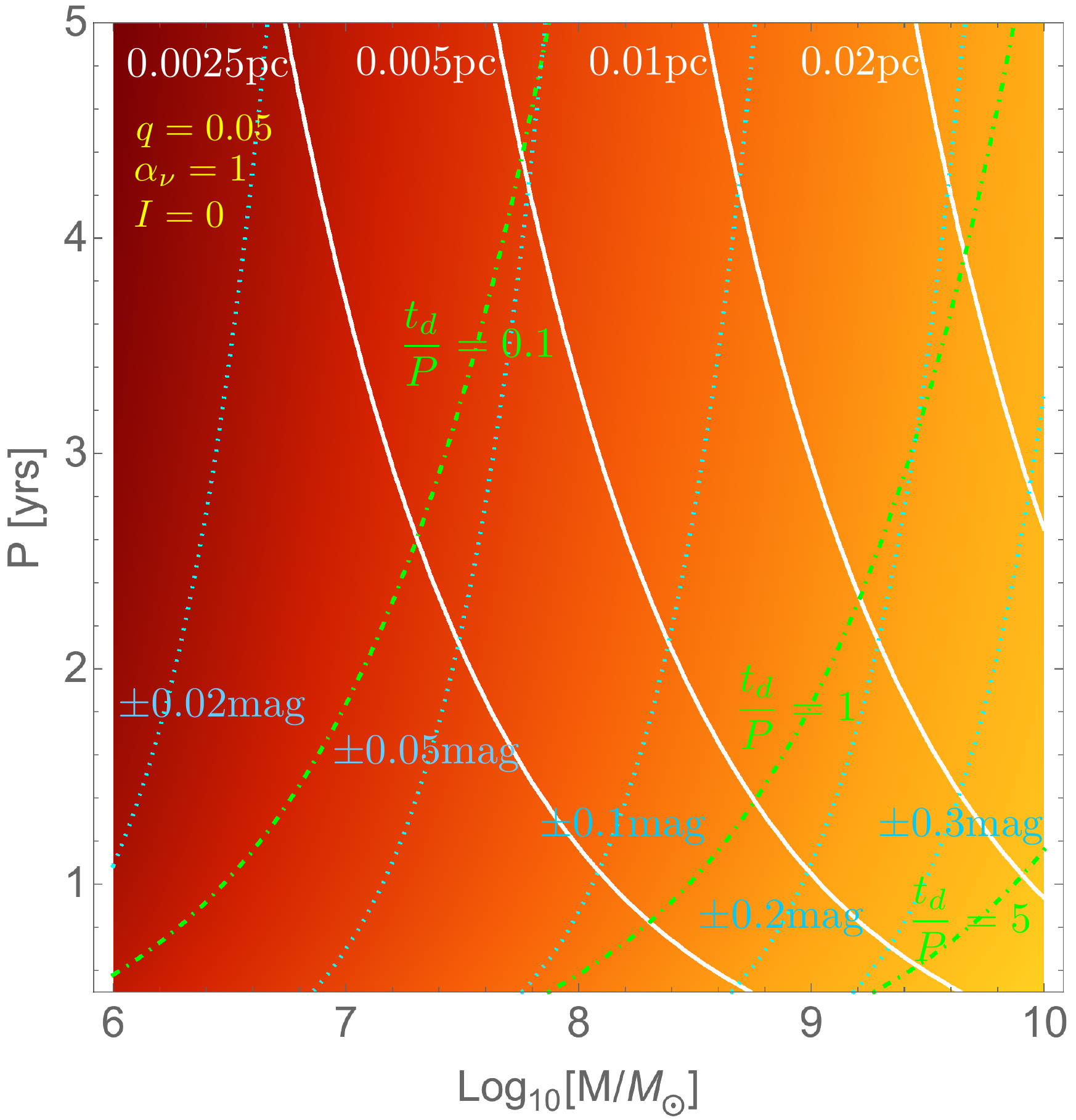} \hspace{20pt} 
\end{array}$
\vspace{-10pt}
\end{center}
\caption{
Representative binary masses and orbital periods for which Doppler boosting
causes significant variability. The orange to yellow color map shows regions of
constant (log) Doppler modulation amplitude from low to high respectively. Cyan
contours delineate specific values of constant Doppler modulation amplitude in
magnitudes. White curves mark constant binary separations, and green curves
indicate different ratios of the light crossing time at the inner edge of a
dust distribution $t_d$ to the binary orbital period (see \S \ref{S:PDs}). We
have assumed a mass ratio of $q=0.05$, an edge-on binary ($I=0$), and a
spectral index  $\alpha_{\nu}=1$.
  }
\label{Fig:boostParams}
\end{figure}

Figure \ref{Fig:boostParams} illustrates combinations of MBHB periods
and masses for which the secondary's orbital velocity causes a
significant modulation in the observed light curve.  The figure
assumes $\alpha_{\nu}=1$, an edge-on view of the binary, and a mass
ratio of $q=0.05$, and shows that MBHBs with orbital periods of a few
years and total masses of $\geq 10^8 \Msun$ can exhibit $\geq 0.1$ mag
modulations due to Doppler boosting.

\section{Calculations of the Dust Echo}
\label{S:Derivation}
\subsection{Reverberation of isotropically variable emission}
\label{S:FISOderivation}

\subsubsection{Spherical dust shell}

We start with the simplest case of a hollow spherical shell of dust with the source
located at its center. We adopt spherical coordinates ($r,\theta,\phi$) with
the source located at the origin, the observer at $(d, \pi/2, 0)$ and the
shell at $(\Rin, \theta, \phi)$.  Assuming that the dust is in radiative
equilibrium with the heating source, we find the dust temperature as a
function of time by equating the power absorbed by a dust grain to that
radiated,
\begin{eqnarray}
\pi a^2_{\eff}\bar{Q}^{\rm src} F^{\rm iso}(t, R_d) = 4\pi a^2_{\rm eff} \int^{\infty}_{0}{Q_{\nu} \pi B_{\nu}\left[T_d(t)\right] \ d \nu }.
\label{Eq:TdISO}
\end{eqnarray}
Here $a_{\rm eff}$ is the effective (spherical) grain radius which describes
the dust cross section for absorption and also the surface area for emission,
$\pi B_{\nu}$ is the blackbody flux from a uniformly emitting dust grain at
temperature $T_d$, $Q_{\nu}$, is the efficiency of dust absorption/emission
and  $\bar{Q}^{\rm src}$ is its average over the source spectrum.

Radiation with wavelength $\lambda \lsim 2 \pi a_{\eff}$ is absorbed
efficiently by dust grains. For longer wavelength radiation, grains of the
same size become transparent.  Hence for the absorption/emission efficiency we
choose $Q_{\nu}=1$ for frequencies above a cutoff $\nu_0 \sim c (2 \pi
a_{\eff})^{-1}$ and a power law fall off in efficiency for lower frequency
radiation \citep[\textit{e.g.}][]{Barvainis:1987}, $Q_{\nu} \equiv
\rm{min}\left[ (\nu/\nu_0)^k, 1\right]$ where $k \geq 0$. \footnote{ For this
form of $Q_{\nu}$, the right-hand side of Eq. (\ref{Eq:TdISO}) can be written
in terms of polylogarithmic functions.} \cite{DraineLee:1984} fit $k=1.6$ for
graphite grains. Because the efficiency for absorption is unity for high
frequency radiation, above $\sim$ $1\mu$m, we take $\bar{Q}^{\rm src} = 1$
throughout.

\begin{figure}
\begin{center}$
\begin{array}{c }
\includegraphics[scale=0.22]{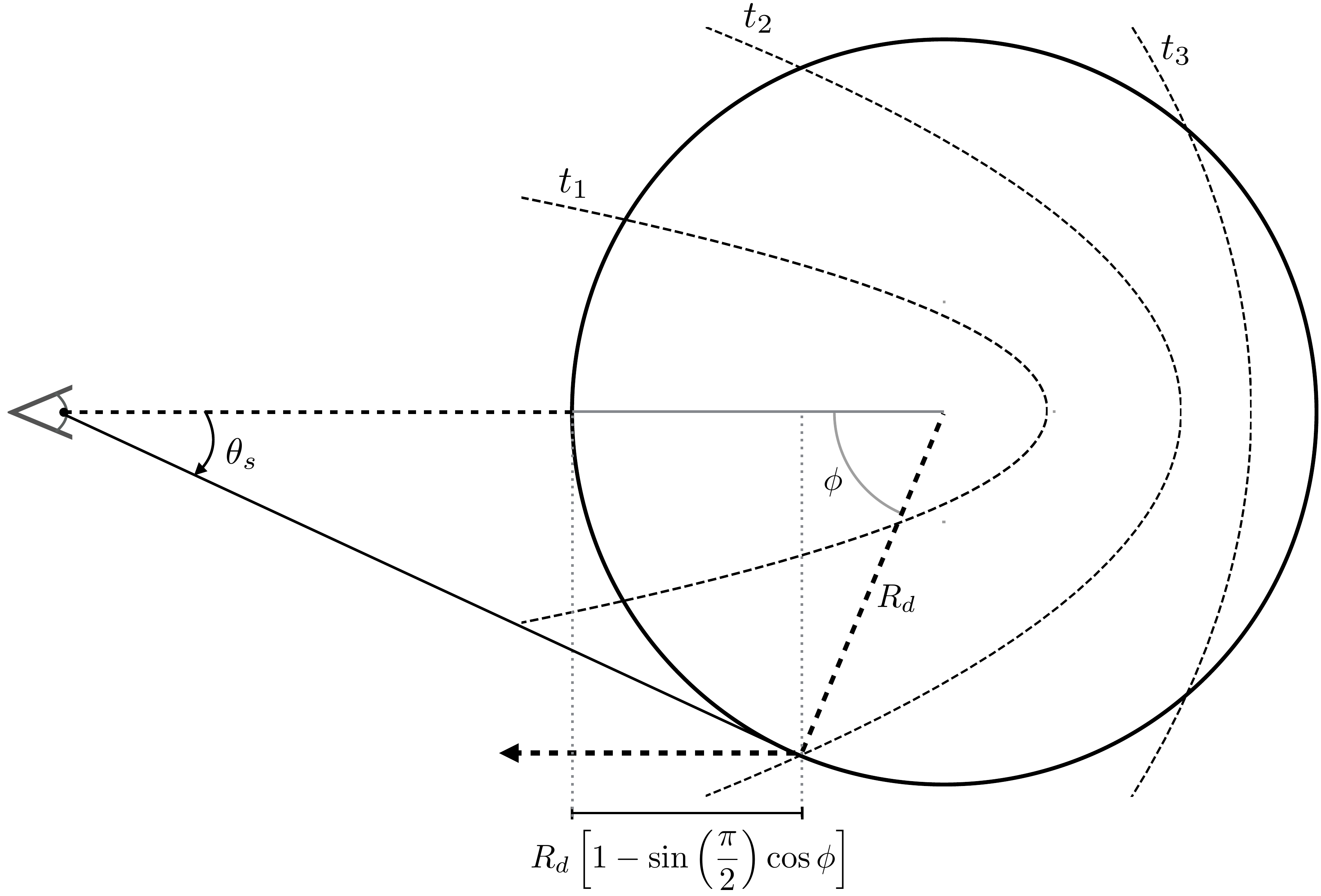} \hspace{20pt} \\ 
 \includegraphics[scale=0.22]{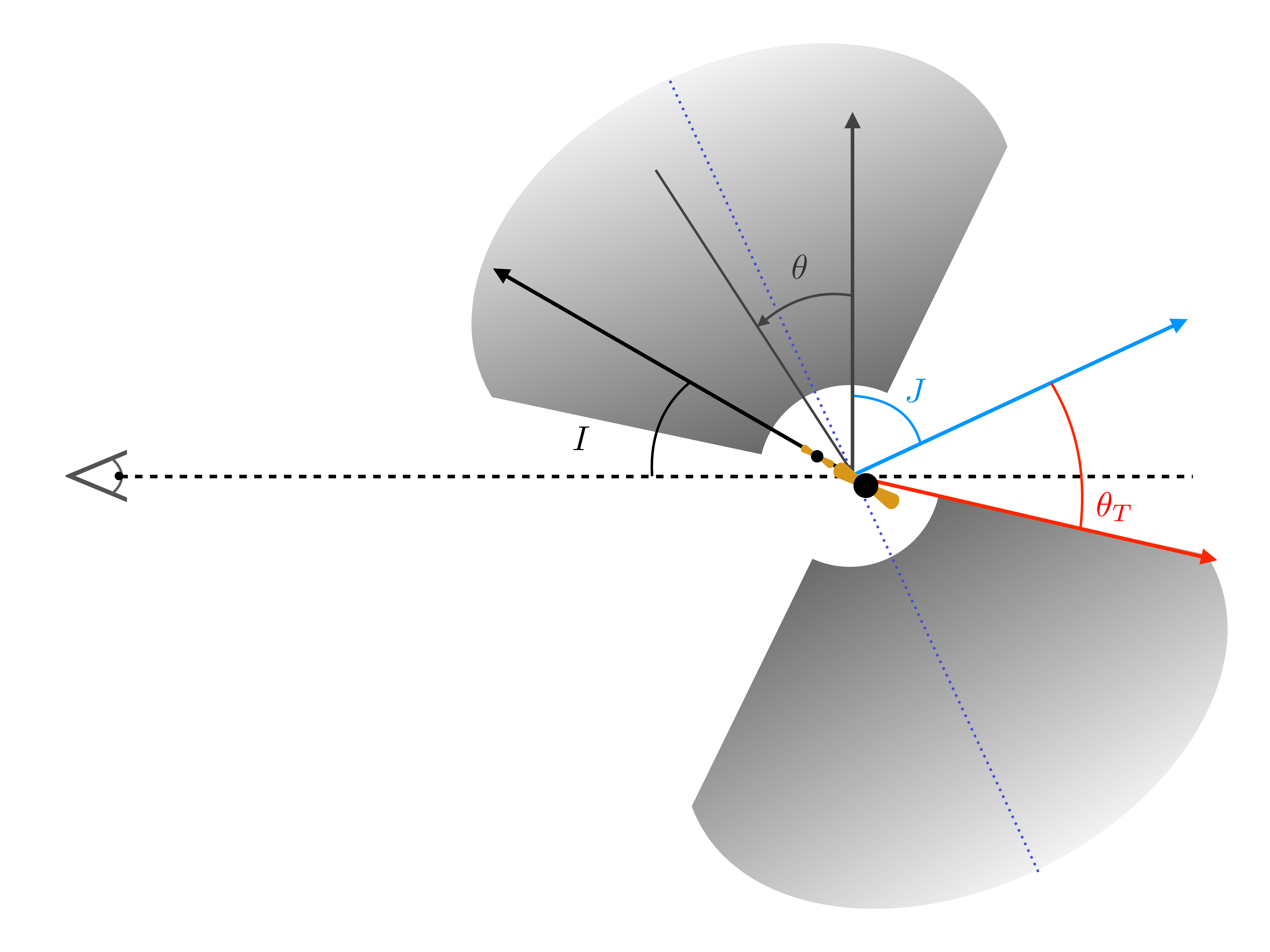} 
\end{array}$
\end{center}
\caption{
{\it Top panel:} light travel time geometry. The circle shows a cross section
of the spherical dust shell. Light leaving the intersection of the dust and
the  ellipse $t_1$ reaches the observer (on the left, at coordinates
$r=d,\theta=\pi/2, \phi=0$) before light leaving the intersections of the
source with $t_2$ and $t_3$. For continuously illuminated dust, the observer
measures a flux summed over all past parabolas intersecting the circle, each
at its own retarded time. {\it Bottom panel:} angles defining the torus
geometry. $I$ is the inclination of binary's orbital plane to the line of
sight, $J$ is the inclination of the symmetry axis of the torus to the plane
perpendicular to the line of sight, $\theta_T$ is the opening angle of the
torus, and $\theta$ is the spherical polar angle in our coordinate system.
}
\label{Fig:Schm}
\end{figure}

The observed flux from a single dust grain at temperature T is
\begin{eqnarray}
F^{\rm{grain}}_{\nu} &=& 2 \pi \int^{\theta_c}_0{ Q_{\nu} B_{\nu}[T_d] \cos{\theta_s} \sin{\theta_s} d \theta_s} 
\\ \nonumber 
&=& \left(\frac{ a_{\rm{eff}} }{d}\right)^2 Q_{\nu} \pi B_{\nu}[T_d], \quad
\theta_c \equiv \sin^{-1}\left( \frac{ a_{\rm{eff}}}{d}\right) .
\end{eqnarray}
where $\theta_s = \theta_c$ is the angle subtended on the sky by a
grain with radius $ a_{\rm{eff}}$ at a distance $d$ from the
observer. From the grain number density and the time-dependent dust
temperature everywhere in the shell (Eq.~\ref{Eq:TdISO}), we compute
the total observed specific flux summing over all dust grains as
\begin{eqnarray}
\label{Eq:FnuSS}
 F_{\nu}(t) &=& \left(\frac{ a_{\rm{eff}} }{d}\right)^2 \int^{2 \pi}_{0}{\int^{\pi}_{0}{ \Sigma_d Q_{\nu}  \pi B_{\nu}\left[T_d(t_{\em})\right]  R^2_d \sin{\theta} \ d \theta d\phi }}   \nonumber   \\
t_{\em}& \equiv & t - \frac{\Rin}{c} \left( 1 - \sin{\theta} \cos{\phi}\right).
\end{eqnarray}
Here $\Sigma_d$ is the surface number density of the dust shell; $\Sigma_d
\rightarrow \pi^{-1} a^{-2}_{\eff}$ in the assumed limit that all optical/UV
radiation is absorbed by the sphere. We have evaluated the temperature at the
retarded time $t_{\em}$; light emitted at the front of the dust shell at time
$t$ and from the location ($\Rin$, $\theta$, $\phi$) at the earlier time
$t_{\em}$ reach the observer at the same time at $t+(d-\Rin)/c$. The top panel
of Figure \ref{Fig:Schm} illustrates this by drawing cross sections of the
ellipsoids of constant light travel time (described by the equation for
$t_{\em}$).

The total flux observed by an instrument with bandpass function $W(\nu)$ is
 \begin{equation}
F_{W}(t) = \int^{\infty}_{0}{  W(\nu) F_{\nu}(t) d \nu  } \sim \int^{\nu_{\rm max}}_{\nu_{\rm min}}{ F_{\nu}(t) d \nu}
\label{Eq:FISO}
 \end{equation}
where we assume for simplicity that $W(\nu)$ is a top-hat function with frequency
limits $\nu_{\rm min}$ and $\nu_{\rm max}$.

\subsubsection{Torus shell}

The spherical dust shell of the previous section is described by a single
geometric parameter, its radius $R_d$. We expand upon the spherical model by
cutting out portions of the sphere to make a torus (or disc) which emits only
from its inner edge. This torus model introduces a second and a third dust
geometry parameter: the opening angle of the torus $\theta_T$ and the
inclination of the torus to the plane perpendicular to the line of sight $J$.
(Note that the opening angle sets the geometrical dust covering
fraction $\cos{\theta_T}$.) These angles and the binary's orbital inclination
angle $I$ are drawn schematically  in the bottom panel of Figure
\ref{Fig:Schm}. To obtain the torus model, we simply set the emission to zero
in the empty cone $J\pm\theta_T$ (and  $[J+\pi]\pm\theta_T$) as shown in this
illustration.

\begin{table*}
%
%
\begin{tabular}{l|l|l|l}
\hline
\hline
  Parameter         & Meaning     & Fiducial Value   &   Notes \\
\hline
\hline 
Source Parameters &  & \\
\hline
\hline 
$L^0$ \quad    & Bolometric luminosity                 & $7\times10^{46}$  erg s$^{-1}$    & Value inferred for PG~1302-102\\
$BC$ \quad    & Bolometric correction from V band                 & $9$    & Typical value for quasars\\
$P$  \quad     & Variability period                           & $R_d /c$  & Match light crossing time of torus\\
\\[-0.2cm]
\multicolumn{4}{|c|}{\it --- --- Additional source parameters needed only in Doppler case: --- --- }\\
\\[-0.2cm]
$\alpha_{\nu}$     & ``in-band'' Spectral index           &  1  & Approximate PG~1302-102 value in optical \\
$\alpha_{\tot}$     & ``bolometric'' Spectral index           &  -1  & Effective spectral index when integrating \\
& & & over entire spectrum (Eq. \ref{Eq:DopAll})\\
$\beta$             & Orbital velocity$/c$                      & $0.07$   & Approximate PG~1302-102 value for fiducial $\alpha_{\nu}$  \\
$I$                 & Orbital inclination                         &  $0$  & 0=edge-on; $\pi/2$=face-on \\
\hline
\hline   
Dust Parameters &  & \\
\hline
\hline 
 $\Rin$               &  Inner edge of dust                    & $1.0$ pc      & Approximate sublimation radius for PG~1302-102 \\
$k$                  & Absorption/emission efficiency exponent           & 1.6                  & Approximate value for graphites \citep{DraineLee:1984} \\
$\nu_0$              & Efficiency cutoff frequency                       & $c/(2 \pi a_{\rm{eff}})$    & Needed for efficiency $Q_{\nu} = \rm{min}\left[ \left( \nu/ \nu_0 \right)^{k}, 1\right]$    \\
$a_{\rm{eff}}$  &  Grain size                                             &  $0.16$ $\mu$m             &  Set by $\nu_0$       \\
\\[-0.2cm]
\multicolumn{4}{|c|}{\it --- --- Additional dust parameters needed only for torus case: --- --- }\\
\\[-0.2cm]
{$J$}                &  Inclination                                & $\pi/2$                     &   0=edge-on; $\pi/2$=face-on \\
{$\theta_T$}         &  Opening angle                               & $\pi/4$                     &   0=sphere ($\cos{\theta_T}$=covering fraction)\\
\end{tabular}
 %
 %
\caption{Parameters of the model and their fiducial values used in dust echo calculations unless stated otherwise.}
\label{Table:params}
\end{table*}

\subsection{The Lighthouse: anisotropic, Doppler-boosted emission} 
\label{S:DopDer}

To compute the effects of Doppler boost on the IR light curve we need to
change the form of the source (optical/UV) flux. We use Eq. (\ref{Eq:Dop1})
for this purpose, with one important distinction: here the line-of-sight
velocity $v_{||}$ is the line-of-sight speed of the secondary BH as observed
by a dust grain at unit position $\mathbf{\hat{r}}_{\rm{dust}}$ in the dust
shell. Written in  barycentric coordinates $(r,\theta, \phi)$,
\begin{eqnarray}
\label{Eq:vlos}
\frac{v_{||}}{c} &=& \frac{\mathbf{v_{s}} \cdot \mathbf{\hat{r}}_{\rm{dust}} }{c} \\ 
& = & \beta  \left[ \cos{I} \cos{(\phi_0+\Omega t)} \sin{\theta}\cos{\phi} \right. \nonumber \\
& + &  \left. \sin{(\phi_0+\Omega t)} \sin{\theta}\sin{\phi} \nonumber \right. \\
& + & \left. \sin{I} \cos{(\phi_0+\Omega t)} \cos{\theta}  \right], \nonumber 
\end{eqnarray}
where for brevity of notation we have written the binary period in
terms of the angular orbital frequency $\Omega = 2 \pi/P$, $\phi_0$ is
the $\phi$ coordinate of the secondary at the reference time $t=0$, and
we have parameterized the secondary's orbital velocity as $\beta \equiv
a (1+q)^{-1} \Omega/c$, which depends on the binary mass ratio $q$, total
mass $M$, and period $P$ through the binary orbital frequency $\Omega$
and separation $a$.

Because the emission is anisotropic, the dust temperature varies across
the surface of the shell, and the analogue of Eq. (\ref{Eq:TdISO}) becomes
\begin{equation}
\label{Eq:TdDop1}
\bar{Q}^{\rm{src}} \int^{\infty}_{0}{F^{\rm{Dop}}_{\nu}(t, R_d, \theta, \phi)  \ d \nu } =   4 \int^{\infty}_{0}{Q_{\nu} \pi B_{\nu}\left[T_d(t,\theta, \phi)\right] \ d\nu}.
\end{equation}

To evaluate the LHS, we emphasize the assumption $\bar{Q}^{\rm{src}}_{\nu} = 1$, that the
dust absorbs all of the light emitted by the central source, or practically,
that the majority of source emission is in the optical and UV. Then we may use
Eq. (\ref{Eq:DopAll}) to write
\begin{equation}
\label{Eq:TdDop}
\left[D(t, \theta, \phi)\right]^{4} \frac{L^0}{4 \pi R^2_d}  =   4 \int^{\infty}_{0}{Q_{\nu} \pi B_{\nu}\left[T_d(t,\theta, \phi)\right] \ d\nu}, 
\end{equation}
where $L^0$ is the bolometric source luminosity and our result is independent
of the shape of the source spectrum. Hence, the Doppler case introduces only two
new source parameters $\beta,$ and $I$ (taking the place of $A$ in the
isotropic case). Once the dust temperature is found from these implicit
equations, it can be used to find $F_{\nu}$ in the same way as derived for an
isotropic source in \S\ref{S:FISOderivation} (in either the spherical shell
or the torus geometry).

\subsection{Model Parameters}

\label{subsec:ModelParams}

The parameters of our model and their fiducial values are summarized in Table
\ref{Table:params} for the convenience of the reader. These fiducial values
are used in calculating the dust echo light curves below, unless stated in the
text otherwise. The parameters are divided into two categories, describing the
central MBHB source and the dust torus, respectively.

\section{Analysis}
\label{S:PDs}

We now identify the effect of the model parameters (see Table
\ref{Table:params}) on the IR light curves. For comparisons of the optical/UV
continuum light curves with their reverberated IR counterparts, we focus on
the average brightness, phase, and variability amplitude of the periodic IR
emission relative to the optical/UV continuum.

In the general case, we compute IR light curves by numerically solving for the
dust temperature in Eqs. (\ref{Eq:TdISO}) or (\ref{Eq:TdDop}) and then
evaluating Eq.~(\ref{Eq:FISO}) for the total in-band flux. We begin by
building up intuition for reverberation of periodic sources by analytically
evaluating a few simplified cases.

\subsection{Spherical dust shell} 
\label{S:Interp:Sphere} 

In the case of an optically thick dust sphere, none of the optical/UV
continuum reaches the observer, but we begin by looking at this case, as it is
the simplest.  This can be regarded mainly as an academic exercise.  However,
the spherical case can be of practical interest, describing systems for which
only periodic IR emission is observed. It can also be applicable if the dust
sphere is only marginally optically thick, and/or if the dust distribution is
clumpy on small scales, making the dust sphere porous (or if the
sphere has a rare narrow `hole' toward the observer).

\subsubsection{Isotropically variable source}
\label{S:Simplestcase}

We first consider isotropic, sinusoidal emission by the central source. The IR
luminosity evaluated at the retarded time $t_{\em}$ (Eq.~\ref{Eq:FnuSS}),
integrated over all frequencies, becomes
\begin{eqnarray}
&&L^{\rm{Iso}}_{\rm{IR}}(t)
=\Sigma_d 4 \pi a^2_{\eff} \int^{2 \pi}_0{\int^{\pi}_0{  \frac{L\left( t_{\em}(t, \theta, \phi) \right)}{16 \pi R^2_d } R^2_d \sin{\theta} d\theta d\phi }} \nonumber \\
&&=  \Sigma_d \pi  a^2_{\eff} L^0  \left[ 
  1 +  A \rm{sinc}{\left( \Omega t_d \right)}  \sin{ \left( \Omega \left( t - t_d \right) \right)  }
   \right].
   \label{Eq:LIR_sp}
\end{eqnarray} 
Here $\rm{sinc}(x) = \sin{x}/x$ is the cardinal sine function, and we
assumed the optical/UV luminosity of Eq.~(\ref{Eq:Fsrc_ISO}),
\begin{equation}
L(t) =  L^0  \left[ 
  1 +  A \  \sin{ \left( \Omega t  \right)  }
   \right].
\end{equation}
with average luminosity $L^0$ and modulation amplitude $A$.\footnote{This
analytic result can be generalized to arbitrary periodic functions by
replacing $L(t)$ with a Fourier series expansion.}

This simple expression in Eq.~(\ref{Eq:LIR_sp}) gives some insight
into the reverberation of a periodic continuum. We find, as expected,
that the average luminosity in the IR is the same as in the
optical/UV. A novel result is that the amplitude of modulation is
diminished by a factor of
\begin{eqnarray}
\frac{A^{\rm{Iso}}_{\rm{IR}}}{A} = \frac{1}{2 \pi}\frac{P}{t_d} \sin{\left[ 2 \pi\frac{t_d}{P} \right]},
\label{Eq:AIRoAUV}
\end{eqnarray}
where $A_{\rm{IR}}$ and $A$ are the amplitudes of the IR and
optical/UV modulations, respectively, and we have defined the
light-crossing time across the dust torus as $t_d \equiv
R_d/c$. Additionally, the IR is modulated at the same period as the
optical/UV continuum, but with a phase lag given by
\begin{eqnarray}
\Phi_{\rm{Iso}} &=& \frac{t_d}{P} -  \left[1 - \rm{sign}\left(\frac{A^{\rm{Iso}}_{\rm{IR}}}{A}\right) \right] \frac{1}{4}  \quad \rm{cycles}
\label{Eq:ISOLags}
\end{eqnarray}
written in fractions of a cycle. When the IR fluctuation amplitude
(Eq.~\ref{Eq:AIRoAUV}) is positive, we recover the expected phase lag
given by the light travel time from the central source to the dust
shell. However, if $A_{\rm{IR}} < 0$, there is an additional
half-cycle phase change. We stress that this half-cycle phase shift
occurs only for optical/UV light-curves that are reflection symmetric
about their average.

\begin{figure}
\begin{center}
\includegraphics[scale=0.4]{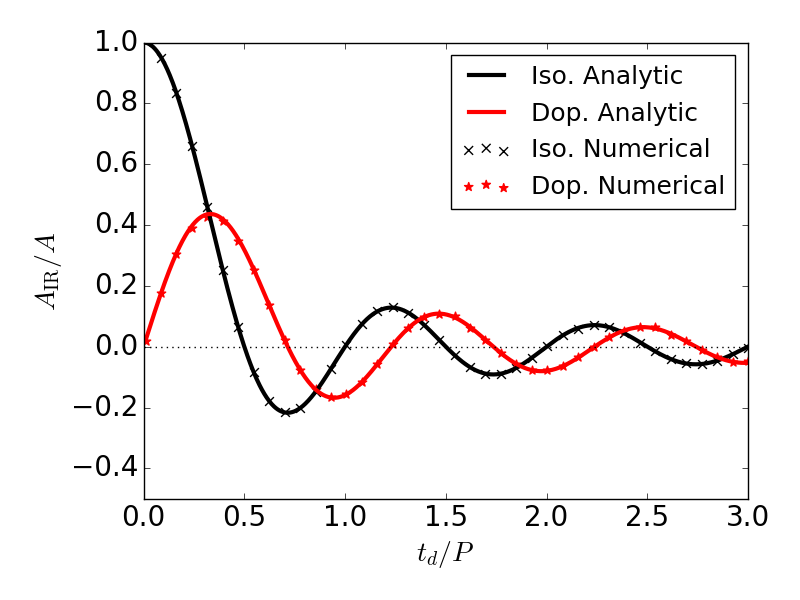}
\end{center}
\vspace{-20pt}
\caption{ The fractional amplitude of IR variability $A_{\rm{IR}} =
  \Delta L_{\rm{IR}}/L_{\rm{IR}}$ relative to the optical/UV amplitude
  $A=\Delta L/L$ for a spherical dust shell which absorbs all
  optical/UV radiation and re-emits it all in IR. The IR amplitude is
  given by the absolute value of the plotted quantity while positive
  and negative values denote a half cycle phase difference. Numerical
  values for both isotropic (black crosses) and Doppler-boosted (red
  stars) sources are computed from the peaks and troughs of solutions
  for IR light curves laid out in \S \ref{S:Derivation}. The analytic
  solutions (solid lines) are Eq. (\ref{Eq:AIRoAUV}) for the
  isotropically varying source (black) and Eq. (\ref{Eq:DopAIR_sp})
  for the specific case of a Doppler source with $\alpha_{\tot}=4$ and
  $v/c \ll 1$.  }
\label{Fig:AIRoAUV_sp}
\end{figure}

Eq. (\ref{Eq:AIRoAUV}) shows that the IR modulation amplitude is
determined by the ratio $t_d/P$. For $t_d/P \rightarrow 0$, the light
travel time across the dust shell is insignificant and the IR
light-curve tracks the optical/UV with the same amplitude.
As $t_d/P$ increases, the ratio $A_{\rm IR}/A$ decreases, falling to
zero for $t_d/P \gg 1$, and whenever the light-crossing time of the
dust shell is an integer multiple of the variability period, where
$t_d/P = \frac{m}{2}$. The analytic result from Eq.~(\ref{Eq:AIRoAUV})
is shown in Figure~\ref{Fig:AIRoAUV_sp} (black curve), along with the
numerical results based on the equations in \S \ref{S:Derivation}
(black crosses).

\subsubsection{Doppler-modulated source} 
\label{S:DopAnyl}

Recall from \S\ref{S:DopDer}, that the dust sees a source flux modulated by
the Doppler factor to the $4^{\rm{th}}$ power, \textit{i.e.},
$\alpha_{\tot}= -1$.  Unfortunately, for this case, we are unable to
find an exact analytic expression for the IR echo.  However, we \textit{do}
find an exact analytic solution when $\alpha_{\tot}=4$, and we find that this
exact solution is a good approximation to the better justified
$\alpha_{\tot}=-1$ case (the two cases converge in the limit that the binary
and the dust torus are face on). Here we present the exact $\alpha_{\tot}=4$
solution, and compare it to the numerical solutions for $\alpha_{\tot}=-1$ --
justifying the use of the analytic $\alpha_{\tot}=4$ results.

When $\alpha_{\tot}=4$, the exact solution for the Doppler IR echo is
\begin{eqnarray}
\label{Eq:LIR_Dop_sp}
L^{\rm{Dop}}_{\rm{IR}} &=&  \Sigma_d \pi  a^2_{\eff} L^0 \times \\ \nonumber 
&\times& \left\{1 +  \gamma \beta \cos{I} \left( \frac{ \sin{\Omega t_d} }{\Omega^2 t^2_d} - 
\frac{\cos{\Omega t_d}}{\Omega t_d}  
\right) 
\cos\left[ \Omega \left(t - t_d\right)\right]
   \right\},
\end{eqnarray}
and hereafter we assume that the binary orbital velocity is only mildly
relativistic and allow the Lorentz factor $\gamma \rightarrow 1$.

The reverberated IR variation follows $\cos\left[\Omega
  \left(t-t_d\right)\right]$, in contrast to the optical/UV continuum
(Eq.~\ref{Eq:vlos}) which follows $\sin\left( \Omega t\right)$.  This
means that the IR echo has a phase lag of
\begin{equation}
\Phi_{\rm{Dop}} = \frac{t_d}{P} - \rm{sign}\left(\frac{A^{\rm{Dop}}_{\rm{IR}}}{A} \right) \frac{1}{4} \quad \rm{cycles,}
\label{Eq:DOPLags}
\end{equation}
i.e. it is a quarter cycle out of phase with the isotropic case. This
can be understood in analogy with the isotropic case, where the IR
echo is delayed by $\Delta t=R_d/c$, which is the average light travel
time difference between the front and back of the dust sphere, or
equivalently the travel time between the front and halfway to the back
($\phi = \pi/2$ and $t_{\em} = t-R_d/c$ in Figure
\ref{Fig:Schm}). This is also the case for the Doppler source,
however, the Doppler flux seen by dust at $\phi = \pi/2$ is
additionally one quarter of a cycle out of phase from the Doppler flux
seen by the observer at $\phi = 0$.

The relative amplitudes also differ from the isotropic case,
\begin{equation}
\frac{A^{\rm{Dop}}_{\rm{IR}}}{\beta \cos{I}}  = \frac{ \rm{sinc}{ \left( 2 \pi \frac{t_d}{P} \right)  }    }{2 \pi \frac{t_d}{P}} - \frac{\cos{ \left(2 \pi \frac{t_d}{P}  \right) } }{ 2 \pi \frac{t_d}{P} },
  \label{Eq:DopAIR_sp}
\end{equation}
where we have replaced A of the isotropic case by a Doppler analogue
$\beta \cos{I}$. It is interesting to note that in the $\alpha_{\tot}=4$ case,
$A^{\rm{Dop}}_{\rm{IR}}$ is the negative derivative, with respect to $2 \pi
t_d/P$, of $A^{\rm{Iso}}_{\rm{IR}}$. We are unaware of this being anything
more than a mathematical coincidence.

Figure \ref{Fig:AIRoAUV_sp} compares analytic (Eq.~\ref{Eq:DopAIR_sp}) and
numerical (\S \ref{S:Derivation} with $\alpha_{\tot}=4$) relative 
IR-variability amplitudes for a Doppler-boosted source to those of the isotropic
source. Differently from the isotropic case, the modulation amplitude falls to
zero for $t_d/P \rightarrow 0$. This is because the Doppler-boosted emission
is observer-dependent, and emanating from a steady rest-frame source; in the
limit that light travel time across the dust is insignificant, conservation of
energy requires that the total emission integrated over a sphere does not vary
in time. Observed IR variability only arises because different light travel
times from different parts of the sphere cause the observer to see different
cross sections of the dust sphere, in the plane perpendicular to the observer,
heated at different look-back times (see Fig.~\ref{Fig:Schm}).

The IR echo of the Doppler-boosted source becomes steady ($A_{\rm
  IR}=0$) at discrete values of $t_d/P$ that are offset by
approximately a quarter cycle compared to the isotropic case, and
given by
\begin{equation}
2 \pi \frac{t_d}{P} = \tan{\left(2 \pi \frac{t_d}{P}\right)}.
\label{Eq:DopZeros}
\end{equation}
The non-trivial solutions can be approximated by
\begin{equation}
\left. \frac{t_d}{P} \right|_{\rm{zeros}} \approx  \frac{2m + 1}{4} - \frac{1}{\pi^2 (2m+1)}  \qquad m=1,2,3... .
\end{equation}
which converge to the condition $t_d \approx \left( \frac{m}{2} +
\frac{1}{4}\right) P$ for $t_d \gg P$.  The largest IR amplitudes
occur for
\begin{equation}
\frac{2 (2 \pi t_d/P)}{(2 \pi t_d/P)^2 -2} =  \tan{\left(2 \pi \frac{t_d}{P}\right)},
\label{Eq:DopExtrm}
\end{equation}
with the first three solutions at $t_d/P \approx 0.33, 0.95, 1.46$ and
subsequent solutions approaching $t_d/P = m/2$, $m=4,5,6...$,
coinciding with the zeros of the isotropic case.\footnote{This
  follows because of the derivative relation between the Doppler and
  Isotropic amplitudes (see discussion below Eq.~\ref{Eq:DopAIR_sp}).}

\begin{figure}
\begin{center}
\includegraphics[scale=0.4]{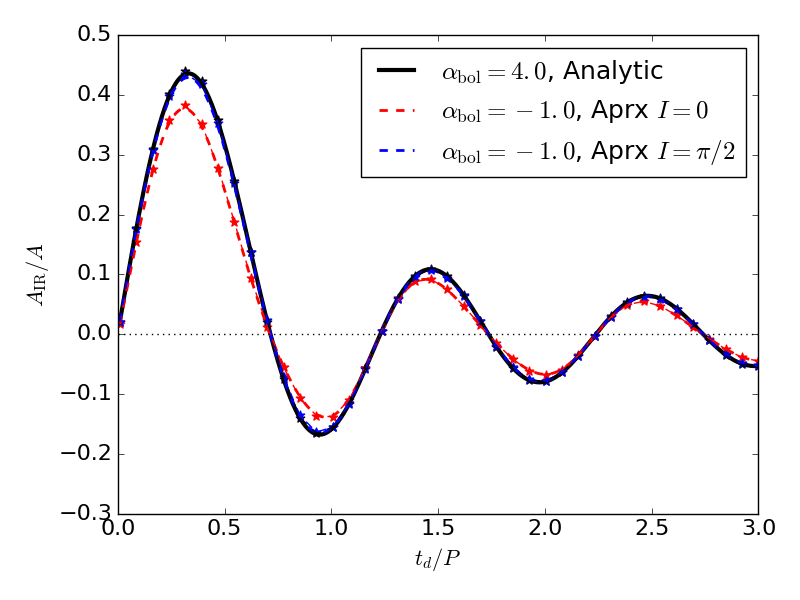}
\end{center}
\vspace{-20pt}
\caption{
  The same as Fig.~\ref{Fig:AIRoAUV_sp}, but showing the difference between
  the physically more relevant $\alpha_{\tot}=-1$ Doppler solutions, and the
  analytical $\alpha_{\tot}=4$ case (black solid curve) that we use to explore
  the properties of Doppler dust echoes. Stars denote numerical evaluations
  for $\alpha_{\tot}=4$ (black), and $\alpha_{\tot}=-1$ ($I=0$ in red and
  $I=\pi/2$ in blue).  For reference, in the latter two cases, we also show an
  approximate, analytic $\alpha_{\tot}=-1$ solution (black and blue dashed
  curves) which are discussed in Appendix~\ref{A:DopApprox}. }
\label{Fig:Dop_alph}
\end{figure}

Finally, in Figure~\ref{Fig:Dop_alph} we show that the properties of
the simple analytic solution discussed above captures the salient
features of the more relevant ``bolometric'' $\alpha_{\tot}=-1$ case.
One difference between the $\alpha_{\tot}=-1$ and $\alpha_{\tot}=4$
solutions is that the former depends on the binary inclination.
The figure shows numerical solutions (as black, red, and blue stars)
for both $\alpha_{\tot}=-1$ and $\alpha_{\tot}=4$.  The exact analytic
solution for $\alpha_{\tot}=4$ is shown by the solid black curve.  For
$\alpha_{\tot}=-1$, we plot the numerical solutions for two extremes, an
edge-on (red) and a face-on (blue) binary. In the case of the
face-on binary ($I=\pi/2$) we find relative amplitudes identical to
the exact $\alpha_{\tot}=4$ solution.  For the edge-on binary ($I=0$)
the solutions for $\alpha_{\tot}=4$ and $\alpha_{\tot}=-1$ predict nearly the same
$t_d/P$ for which zero and maximum relative amplitude occur, but the
simple analytic $\alpha_{\tot}=4$ solution tends to over-predict the
maximum relative amplitudes somewhat. In summary,
Figure~\ref{Fig:Dop_alph} shows that we may use the exact
$\alpha_{\tot}=4$ solutions as a very good estimate of the maximum
relative IR amplitudes from a Doppler-boosted binary.  Finally, Figure
\ref{Fig:Dop_alph} also plots approximate analytic solutions for
$\alpha_{\tot}=-1$ (as dashed blue/red curves; see the Appendix
\ref{A:DopApprox} for details).
\footnote{ While
  it does not affect the analysis of the relative IR amplitude, we
  note that the phases of IR echoes with $\alpha_{\nu}<3$ are one half
  of a cycle out of phase with those with $\alpha_{\nu}>3$. }

\subsubsection{Absorption/emission efficiency}

In the limit of our analytic solutions for the IR luminosity, where we
integrate the dust emission over all frequencies, the dust
absorption/emission efficiency does not affect our result (see
Eqs.~\ref{Eq:TdISO}~and~\ref{Eq:TdDop}). However, when integrating
over a specific IR band, the variation in dust temperature shifts the
dust spectrum blue-ward and red-ward over a variability cycle. This
introduces an additional filter-dependent IR variability, which
depends on the absorption/emission efficiency of the dust --
specifically on the frequency at which the cutoff in efficiency occurs
relative to the observing band.

\begin{figure}
\begin{center}$
\begin{array}{c}
\includegraphics[scale=0.4]{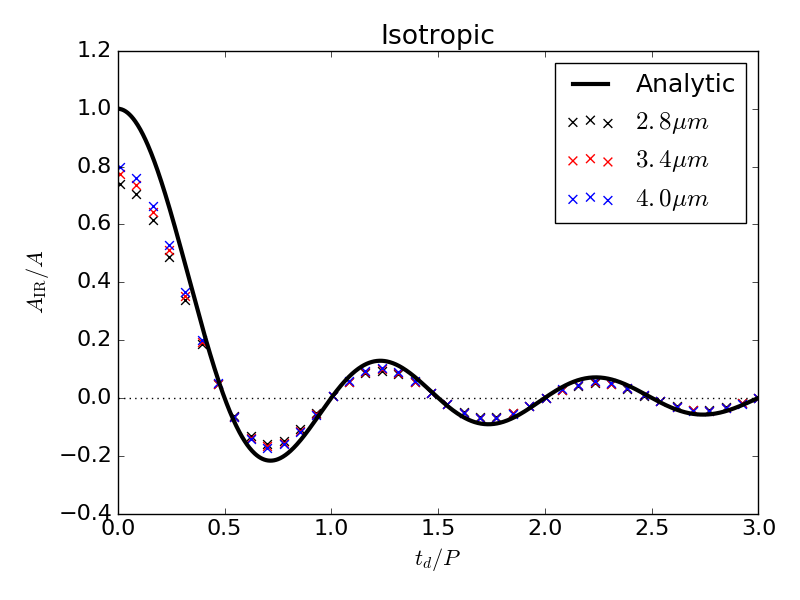} 
\end{array}$
\end{center}
\vspace{-20pt}
\caption{The same as Fig.~\ref{Fig:AIRoAUV_sp} except showing the
  effect of integrating over a finite IR band between $2.8\mu$m and
  $4.0\mu$m (i.e. the WISE W1 band) and allowing for non-unity dust
  absorption/emission efficiency.  
 }
\label{Fig:Qvon}
\end{figure}

Figure \ref{Fig:Qvon} shows the effects of the adopted dust
absorption/emission efficiency and finite frequency band. This figure plots
the analytic solution and compares it with numerical solutions that integrate
over a narrow range of frequencies from $2.8\mu$m to $4.0\mu$m (the WISE W1
band), with different values of the cutoff frequency $\nu_0$ in the
absorption/emission efficiency (see Table \ref{Table:params}). The small
temperature changes are not enough to shift the dust spectrum across the
observing band and boost the IR variability. Instead, we see that the effect
of narrowing the observing band is to decrease the relative IR variability
amplitude. This is likely because hotter dust emits less in the chosen IR
observing band, decreasing the peak IR emission corresponding to when the
central source is brightest, buffering the IR variability amplitude.

\subsubsection{Light Curves}

\begin{figure}
\begin{center}$
\begin{array}{c}
\includegraphics[scale=0.4]{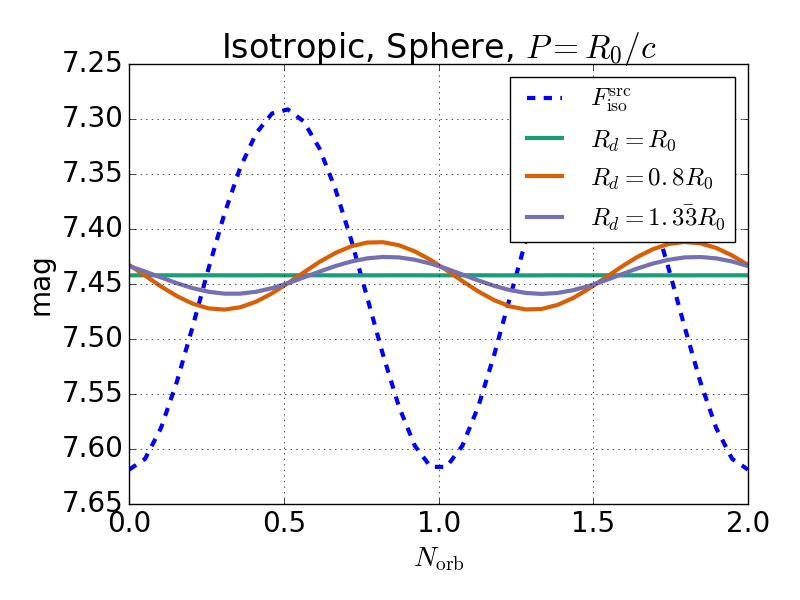}  \vspace{-20pt} \\
\includegraphics[scale=0.4]{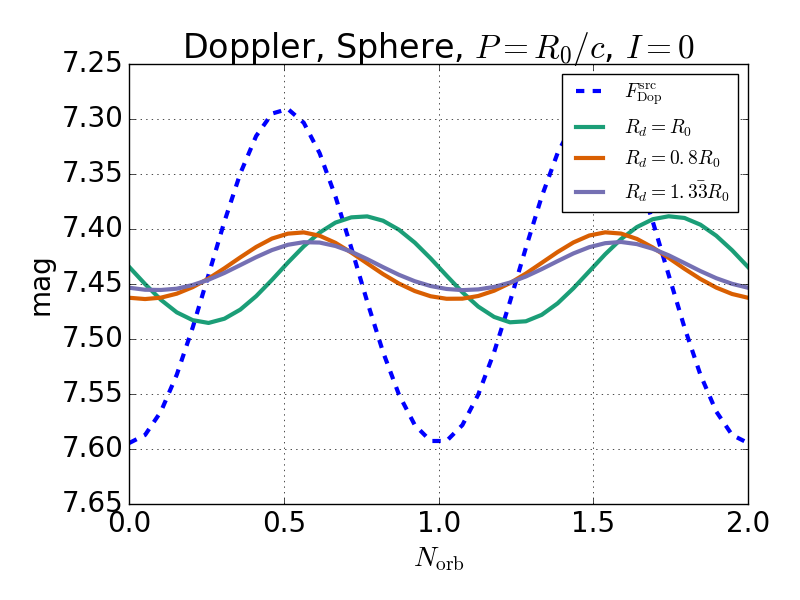} 
\end{array}$
\end{center}
\vspace{-20pt}
\caption{
Spherical dust shell model. The dashed blue curve shows the optical/UV
continuum, and the solid lines show the IR light curves from reverberation
from a spherical dust shell with radius $R_d$ (measured in units of
$R_0=1.0$pc; see Table \ref{Table:params} for other fiducial parameter
values). The top (bottom) panel is for an isotropic (Doppler-boosted) central
source. The magnitude scale is arbitrary.
}
\label{Fig:SphDopvISO}
\end{figure}

Numerically evaluating the expressions in \S \ref{S:Derivation}, we next plot
the optical/UV (source) and IR (echo) light curves for the spherical case in
Figure~\ref{Fig:SphDopvISO}. The top (bottom) panel of this figure assumes an
isotropic (Doppler-boosted) source. We include the dust absorption/emission
efficiency and integrate over all frequencies. To compute the IR Doppler light
curves below, we use the effective 'broad-band' $\alpha_{\tot}=-1$, while for
the optical/UV Doppler light curves, we use the fiducial $\alpha_{\nu}=1$.
This means that to compare relative variability amplitudes from a plot like
Figure~\ref{Fig:SphDopvISO} to the predicted values for the Doppler case, as
in Eq. \ref{Eq:DopAIR_sp}, one must divide the IR amplitude by the optical
amplitude and multiply by a factor of $(3-1)/(3-(-1))=1/2$ -- this factor is a
``bolometric correction'' for the Doppler modulation amplitude from the
optical-only band (measured by Astronomers on Earth) to the full optical/UV
spectrum (seen by the dust).

The model parameters and their
fiducial values are given in Table~\ref{Table:params}.

The IR amplitudes of the light curves in Figure \ref{Fig:SphDopvISO}
are in agreement with the predictions from Figure
\ref{Fig:AIRoAUV_sp}. Each IR light curve is computed for a dust sphere with
radius given in units of $R_0 = 1$pc, as labeled in the figure
legend. For the choice of $P = R_0/c$ this gives $t_d/P = 0.8$ for the
orange curve, $t_d/P = 1$ for green, and $t_d/P = 4/3$ for
purple.

For the isotropic case, we confirm that the IR light curves lag the
optical/UV continuum by the fraction of a cycle given in
Eq. (\ref{Eq:ISOLags}). Recall that for $A_{\rm{IR}}/A$ of different
signs, the corresponding light curves are half a cycle out of
phase. This means that the orange curve, for which $A_{\rm{IR}}/A <
0$, is $0.5+0.8=1.3$ cycles behind the optical/UV (shifted to the
right in Figure \ref{Fig:SphDopvISO}), while the purple curve, for
which $A_{\rm{IR}}/A > 0$, is $4/3$ cycles behind.

Comparison of the top and bottom panels of Figure \ref{Fig:SphDopvISO}
shows the predicted $1/4$-cycle lag between the isotropic and Doppler
IR light curves (accounting for the half-cycle phase shifts discussed
above).

\begin{figure}
\begin{center}
\includegraphics[scale=0.4]{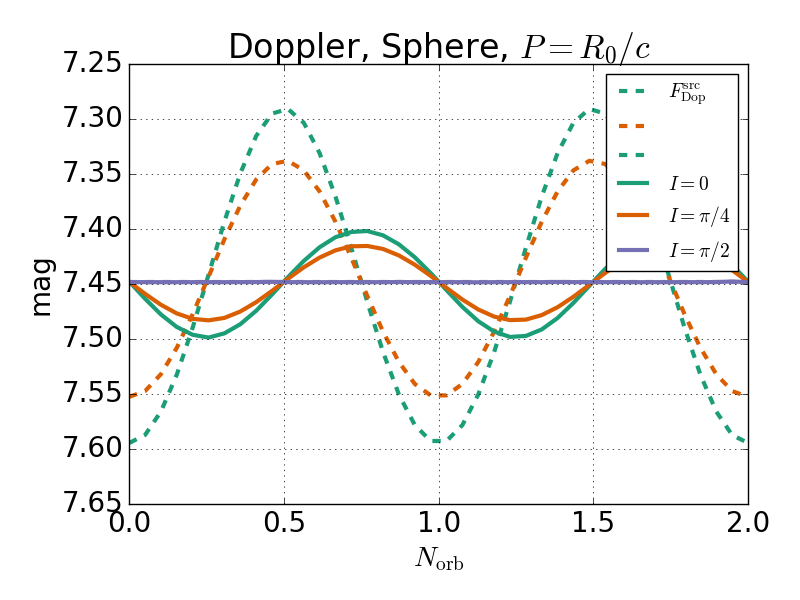}
\end{center}
\vspace{-20pt}
\caption{
The same as the bottom panel of Fig.~\ref{Fig:SphDopvISO} but for fixed $R_d =
R_0$ and three different binary inclination angles ($I$) in the Doppler case
(with $I=0$ corresponding to edge-on and $I=\pi/2$ to face-on; see
Table~\ref{Table:params} for the fiducial parameter choices).
  }
\label{Fig:Sph_VarI}
\end{figure}

Figure~\ref{Fig:Sph_VarI} shows the importance of the binary inclination in
the Doppler case. The optical/UV and IR light curves both exhibit zero
modulation amplitude as the binary approaches a face-on configuration. For the
optical/UV this is because $v_{||} \rightarrow 0$ for a face-on binary. For
the IR emission this is because, once we have averaged over all azimuthal
angles (directions perpendicular to the line of sight), there is no time
variation in dust temperature along the line of sight. Note that there will be
line-of-sight variations in this azimuthally-averaged dust temperature even
for a face-on binary if the dust distribution is not symmetric about the
plane perpendicular to the line of sight and containing the source
(\textit{e.g.} for a misaligned torus; as discussed in the next section).

\begin{figure*}
\begin{center}$
\begin{array}{c c c}
\hspace{-1cm}\includegraphics[scale=0.3]{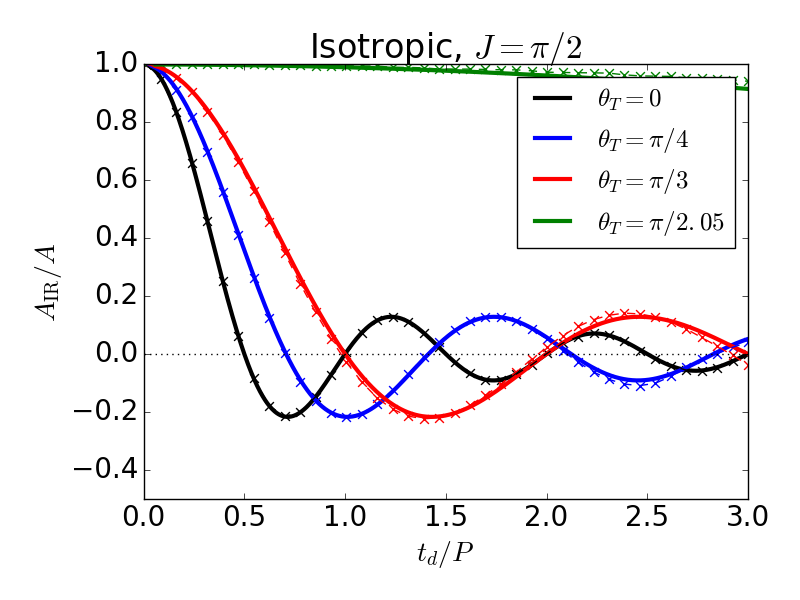}\hspace{-0.0cm} &
\includegraphics[scale=0.3]{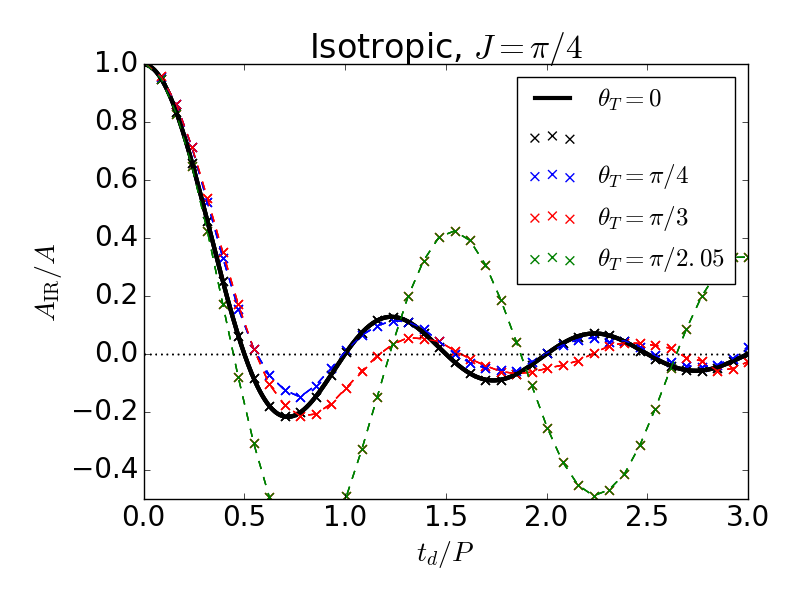}\hspace{-0.0cm} &
\includegraphics[scale=0.3]{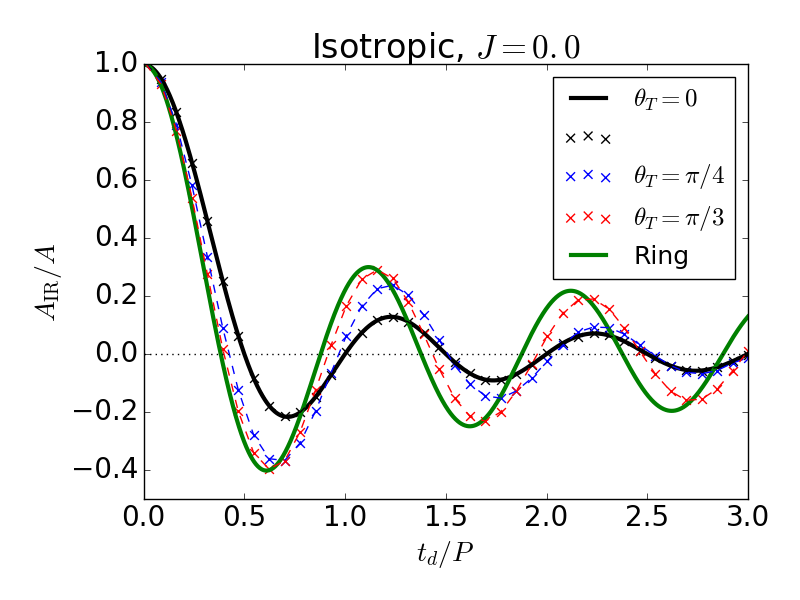} 
\end{array}$
\end{center}
\vspace{-15pt}
\caption{The same as Fig.~\ref{Fig:AIRoAUV_sp} but for a torus
  geometry. Each panel varies the opening angle $\theta_T$ of the dust
  torus for a different torus inclination angle $J$. The solid lines show
  analytic solutions for a face-on ($J=\pi/2$) torus from
  Eq. (\ref{Eq:LIR_thT}). The green line in the right panel is the
  analytic solution Eq. (\ref{Eq:ISO_Ring}) for an edge-on dust ring
  ($J=0$, $\theta_T = \pi/2$). The crosses show results of numerical
  calculations (\S\ref{S:Derivation}).  }
\label{Fig:AIRoAUV_tr}
\end{figure*}

\begin{figure*}
\begin{center}$
\begin{array}{c c}
\hspace{-0.7cm}\includegraphics[scale=0.45]{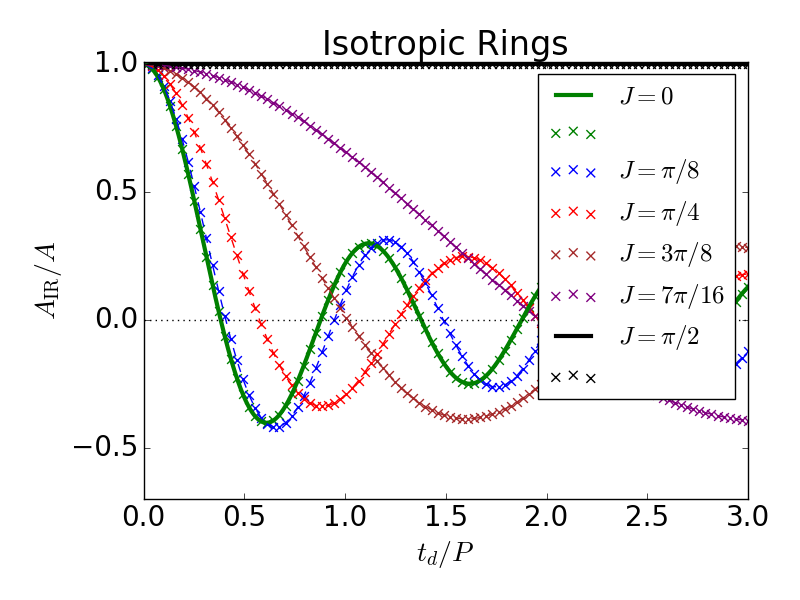} &
\hspace{-0.7cm}\includegraphics[scale=0.45]{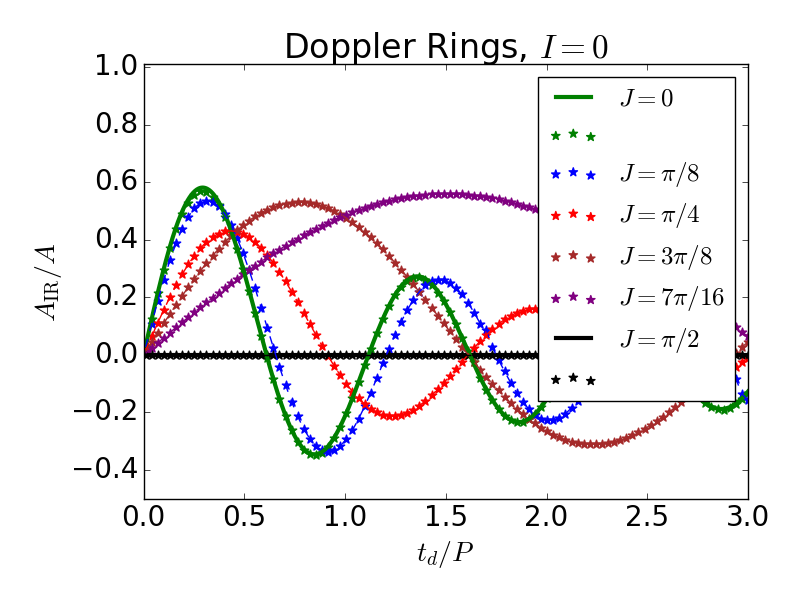} 
\end{array}$
\end{center}
\vspace{-2\baselineskip}
\caption{The relative IR modulation amplitude for thin dust rings
  ($\theta_T \rightarrow \pi/2$) with different inclinations $J$ to
  the line of sight. The left (right) panel is for an isotropically
  pulsating (Doppler boosted) source. }
\label{Fig:AIRoAUV_Rings}
\end{figure*}

\begin{figure*}
\begin{center}$
\begin{array}{c c c}
\hspace{-1cm}\includegraphics[scale=0.3]{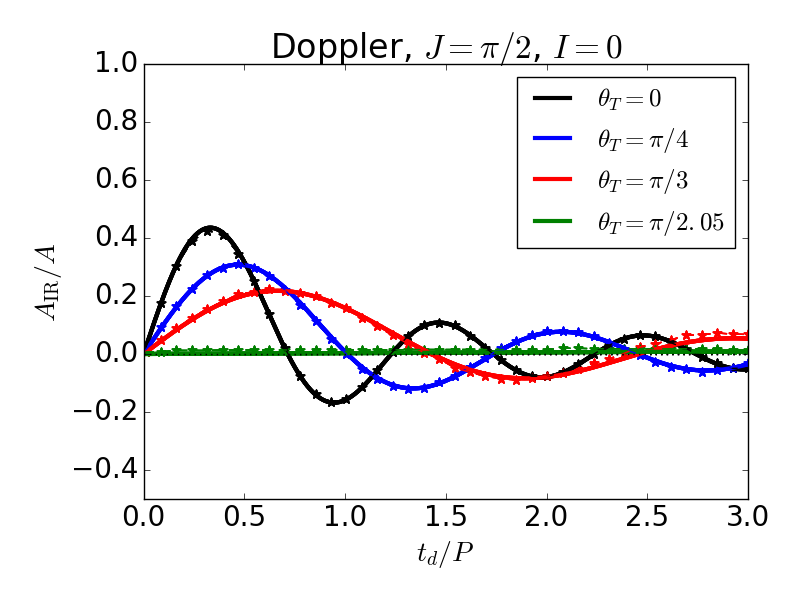}\hspace{-0.0cm} &
\includegraphics[scale=0.3]{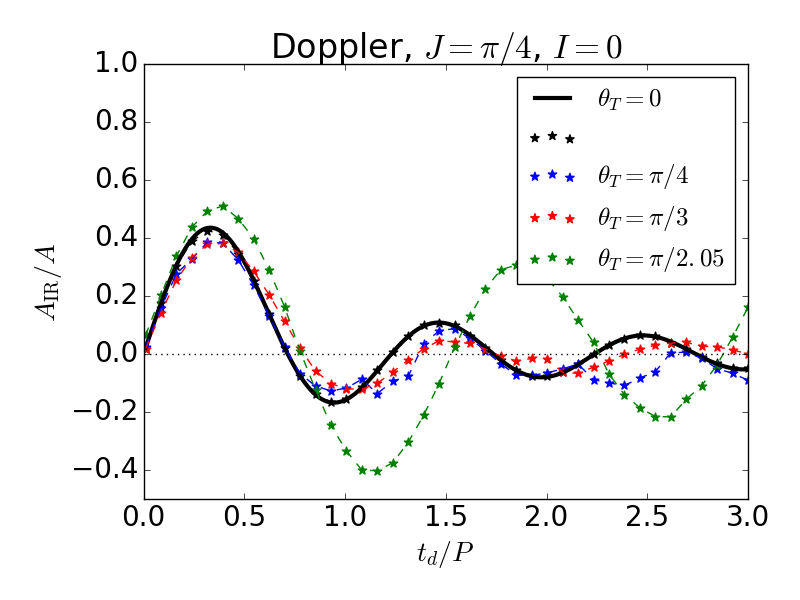}\hspace{-0.0cm} &
\includegraphics[scale=0.3]{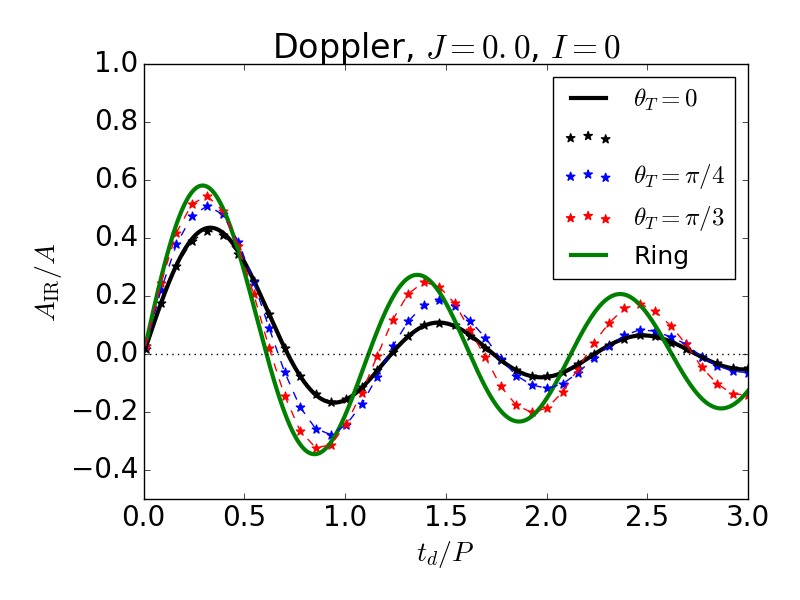} 
\end{array}$
\end{center}
\vspace{-15pt}
\caption{
The same as Figure \ref{Fig:AIRoAUV_tr}, but for a Doppler-boosted source
with an edge-on binary. Solid lines again plot analytic solutions.
}
\label{Fig:AIRoAUVDop_tr}
\end{figure*}

\subsection{Torus} 
\label{S:Interp:ThinTor}

We next extend the results of the previous section to a torus geometry, i.e.
by excising regions of the spherical shell with $\theta \leq \theta_T$.   When
the observer is looking down the axis of the torus ($J=\pi/2$: see Figure
\ref{Fig:Schm}), we find
\begin{eqnarray}
L^{\rm{Iso}}_{\rm{IR}} &=&  \Sigma_d \pi  a^2_{\eff} L^0  \cos{\theta_T} \times  \nonumber \\
&\times& \left\{   1 +  A  \ \rm{sinc}{\left( \Omega t_d \cos{\theta_T}\right)}  \sin{ \left( \Omega \left[t - t_d \right] \right)  }
   \right\} \nonumber
\end{eqnarray}
\begin{eqnarray}
&&L^{\rm{Dop}}_{\rm{IR}} =  \Sigma_d \pi  a^2_{\eff} L^0 \cos{\theta_T} \bigg\{   1 +  \beta \cos{I}   \bigg. \\ \nonumber
 &&\times  \left. \left( \frac{ \rm{sinc}{ \left[ \Omega t_d  \cos{\theta_T} \right]  }    }{\Omega t_d} -  \frac{\cos{ \left[\Omega t_d  \cos{\theta_T} \right]}}{\Omega t_d }  \right) \cos{\left( \Omega \left[t - t_d \right] \right)}
   \right\} ,
   \label{Eq:LIR_thT}
\end{eqnarray}
and the relative amplitudes are 
\begin{eqnarray}
\frac{\Delta L^{\rm{Iso}}_{IR}/(L^0 \cos{\theta_T})}{\Delta L /L^0} \equiv \frac{A^{\rm{Iso}}_{\rm{IR}}}{A} &=&  \rm{sinc}{\left( 2 \pi \frac{t_d}{P} \cos{\theta_T} \right)}  \nonumber 
\end{eqnarray}
\begin{eqnarray}
&&\frac{\Delta L^{\rm{Dop}}_{IR}/(L^0 \cos{\theta_T})}{\Delta L /L^0} \equiv \frac{A^{\rm{Dop}}_{\rm{IR}}}{\beta \cos{I}}  \\ \nonumber
&&=  \frac{ \rm{sinc}{ \left( 2 \pi \frac{t_d}{P} \cos{\theta_T} \right)  }    }{2 \pi \frac{t_d}{P}} - \frac{ \cos{ \left(2 \pi \frac{t_d}{P} \cos{\theta_T} \right) } }{ 2 \pi \frac{t_d}{P} }.
   \label{Eq:AIR_thT}
\end{eqnarray}
Because the dust distribution is centered around the source, the phase
lags are identical to those in the spherical shell case
(Eqs. \ref{Eq:ISOLags} and \ref{Eq:DOPLags}).

\subsubsection{Isotropically variable source}
\label{S:IsoTorAnyl}

Figure \ref{Fig:AIRoAUV_tr} explores the isotropic, torus solutions. The
leftmost panel shows the IR to optical/UV amplitude ratios for different
opening angles, for face on tori (analytic solutions as solid curves and
numerical evaluations shown as crosses). The middle and right panels of Figure
\ref{Fig:AIRoAUV_tr} show numerical solutions for tori inclined by $J=\pi/4$
and for $J=0$ (edge-on).

The effect of increasing the torus opening angle $\theta_T$ is
two-fold. First, it decreases the total IR luminosity -- this is not
relevant for the fractional variability amplitude, but is important
for the absolute IR luminosities. Second, it moves the location of the
zeros of the IR amplitude to larger $t_d/P$. This is because the IR
variability is nullified, in the isotropic case, when an integer
number of variability periods matches the light crossing time along
the line-of-sight dust structure. Depending on the tilt $J$ of the
torus, $\theta_T$ changes this line-of-sight extent, and hence the
values of $t_d/P$ for which $A_{\rm IR}=0$.

Specifically, for a face-on torus ($J=\pi/2$), the $\theta_T$-dependent line-
of-sight extent of the dust shell is $2 R_d\cos{\theta_T}$ (this can be
discerned from Eq. (\ref{Eq:FnuSS}) and visualized in Figure \ref{Fig:Schm}).
As the torus is tilted away from $J=\pi/2$, the relationship between the
closest and furthest points of the sphere along the line of sight becomes less
dependent on $\theta_T$. This can be observed by comparing the left and middle
panels of Figure~\ref{Fig:AIRoAUV_tr}.

To bracket the dependence on $\theta_T$, we consider the extreme cases
of thin dust rings ($\theta_T \rightarrow \pi/2$). These are shown as
green curves in Figure \ref{Fig:AIRoAUV_tr}. Because $\lim_{x \to 0}
\rm{sinc}{(x)} = 1$, Eq. (\ref{Eq:AIRoAUV}) tells us that the limit of
a face-on ring ($\theta_T \rightarrow \pi/2$, $J=\pi/2$), recovers the
optical/UV amplitude at all $t_d/P$, but at lower IR luminosity set by
the covering fraction of the thin ring. This is simply because
light-travel time effects are no longer important for a face-on ring
(left panel of Figure \ref{Fig:AIRoAUV_tr}); as $\theta_T \rightarrow
\pi/2$ the $A_{\rm{IR}}/A$ curves stretch out further to the right at
the expense of lower IR luminosity. Hence for a face-on ($J=\pi/2$)
torus, the limiting behaviour is set by the black curve for a dust
sphere, and a line at $A_{\rm{IR}}/A=1$ for a thin (zero-luminosity)
dust ring.

In the limit of a thin, edge-on ring
($\theta_T \rightarrow \pi/2$, $J=0$) 
the solution for reverberated isotropic emission becomes
\begin{eqnarray}
L^{\rm{Iso}}_{\rm{IR}} &=&  \Sigma_d \pi  a^2_{\eff} \frac{ a_{\rm{eff}} }{ R_d } L^0  \left\{ 1 +  A  \ J_0\left(\Omega t_d\right)  \sin{ \left( \Omega \left[ t - t_d \right] \right)  }    \right\} \nonumber \\
    &&\rm{(Edge-On-Ring)},
   \label{Eq:ISO_Ring}
\end{eqnarray}
where $J_0(z)$ is the zeroth-order Bessel function of the first kind, and the
solution is valid for $a_{\rm{eff}} \ll R_d$. This solution is plotted in the
right $J=0$ panel of Figure \ref{Fig:AIRoAUV_tr} and shows that, for a $J=0$
torus, the possible amplitudes are bracketed again by the two analytic
solutions for a sphere and an edge-on ring. We plot the numerical solution for
a tilted dust ring with $J = \pi/4$ in the middle panel of Figure
\ref{Fig:AIRoAUV_tr}, and find again that the dust ring (which suffers the
least amount of averaging/cancellations) exhibits the largest relative IR
amplitudes.

To investigate further the dependence on the torus inclination angle $J$, the
left panel of Figure~\ref{Fig:AIRoAUV_Rings} shows $A_{\rm{IR}}/A$ for
different inclinations, for the limiting case of a thin dust ring. We find
that as the inclination increases from edge-on ($J=0$) to face-on ($J=\pi/2$),
nearly the full range  $0<A_{\rm{IR}}/A<1$  becomes available for all $t_d/P
\gsim 0.4$. In practice, the viable parameter space will be set by the total
IR luminosity which places a lower limit on the thickness (i.e. covering
fraction) of the dust ring.

\subsubsection{Doppler-modulated source} 
\label{S:DopTorAnyl}

Figure \ref{Fig:AIRoAUVDop_tr} explores the dependence of IR
variability amplitude on dust geometry for a Doppler-boosted source
with an edge-on binary inclination. We find similar $\theta_T$
dependence as in the isotropic case; for the face-on torus (left
panel) the location of $A_{\rm{IR}}/A$ zeros stretches out along the
$t_d/P$ axis as $\theta_T$ increases. However, contrary to the
isotropic case, the limit of a $\theta_T \rightarrow \pi/2$ is
$A_{\rm{IR}}/A=0$, rather than $A_{\rm{IR}}/A=1$. In both cases, this
is a reflection of the intrinsic source behaviour in the limit that
light-travel time effects become negligible, as discussed above.

As for the isotropically varying source, Fig.~\ref{Fig:AIRoAUVDop_tr} shows
the limiting cases of thin dust rings in the Doppler case (green lines). For a Doppler-boosted
source (again using $\alpha_{\tot}=4$ following the discussion in \S
\ref{S:DopAnyl}), the edge-on dust ring produces the IR echo
\begin{eqnarray}
L^{\rm{Dop}}_{\rm{IR}} &=&  \Sigma_d \pi  a^2_{\eff}  \frac{ a_{\rm{eff}} }{ R_d } L^0   \left\{ 1 +  \beta \cos{I}  \ J_1\left( \Omega t_d \right)  \cos{ \left( \Omega \left[ t - t_d \right] \right)  }
   \right\} \nonumber \\ 
   &&\rm{(Edge-On-Ring)},
   \label{Eq:DOP_Ring}
\end{eqnarray}
where $J_1(z)$ is the first-order Bessel function of the first
kind.\footnote{Note that $-A^{\rm{Dop}}_{\rm{IR}}$ is the derivate of
$A^{\rm{Iso}}_{\rm{IR}}$ with respect to $t_d/P$.} This solution is plotted in
the right ($J=0$) panel of Figure \ref{Fig:AIRoAUVDop_tr}. The IR echo for a
dust ring with an inclination of $J=\pi/4$ is shown in the middle panel of the
same figure. Appendix \ref{A:DopApprox} derives approximate edge-on ring
solutions for the $\alpha_{\tot}=-1$ case, finding that the $\alpha_{\tot}=4$
solution here is adequate for discerning the behavior of IR dust echoes.

We explore the dependence on dust inclination further for a Doppler-boosted
source in the right panel of Fig.~\ref{Fig:AIRoAUV_Rings}. As in the isotropic
case, we find that the main effect of increasing $J$ (towards face-on) is to
stretch out $A^{\rm{Dop}}_{\rm{IR}}(t_d/P)$ in $t_d/P$. Figure
\ref{Fig:AIRoAUV_Rings} shows that even the purple $J = 7 \pi/16$ curve
eventually reaches the same maximum relative amplitude as the $J=\pi/2$ curve
($\sim0.6$), but at $t_d/P \sim 1.5$ instead of $t_d/P \sim 0.3$. This is
simply because the closer to face on, the further away the dust ring needs to
be to retain the same difference in light-travel times. Specifically, for a
dust ring tilted by the angle $\delta \phi$ away from face on, the difference
in light-travel time from the bottom {\it vs} the top of the ring is $\delta t
= R_d/c \sin{\Delta \phi}$. So $R_d$ must be made large enough for $t_d/P
\sin{\Delta \phi} \sim 0.3$, the value where the Doppler-boosted IR amplitudes
reach a maximum in the $J=0$ case.

In summary, the largest IR variability amplitudes are realized for the
thinnest, ring-like dust distributions. If the dust ring can be oriented at
any inclination to the line of sight, then for isotropic illumination, the
reverberated IR amplitude can be any fraction ($\leq1$) of the optical/UV
modulation amplitude for $t_d/P \gsim 0.4$, and is restricted to be between
one and a non-zero minimum value for $t_d/P \lsim 0.4$. For the same
orientable dust ring illuminated by an edge-on Doppler-boosted binary source,
the IR variability amplitude has a maximum value of $\approx0.6$ for $t_d/P
\gsim 0.3$, and smaller maximum values below $t_d/P \lsim 0.3$. For a face-on
binary inclination however the relative IR amplitude approaches infinity
because the optical/UV variability drops to zero while the IR variability does
not (see next subsection). The overall IR flux, and/or any other independent
constraint on the thickness, size, and orientation of the dust structure will
restrict the allowed IR variability amplitudes from those presented here for
the thin dust rings.

\subsubsection{Light curves}

\begin{figure}
\begin{center}$
\begin{array}{c c}
\includegraphics[scale=0.2]{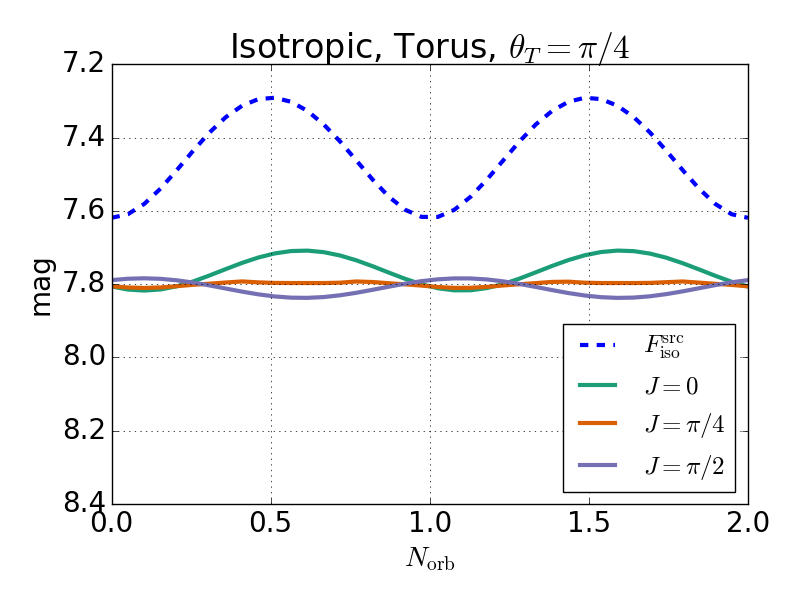}&
\includegraphics[scale=0.2]{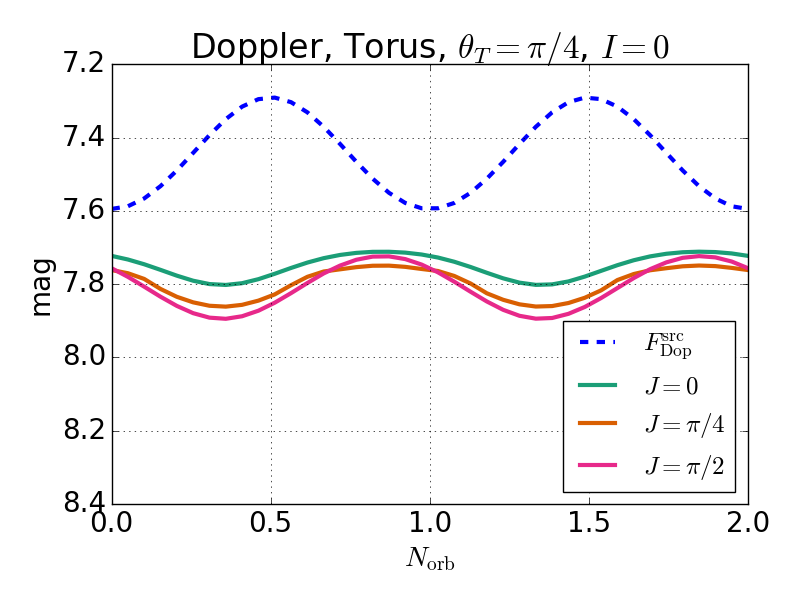} \vspace{-11pt} \\        
\includegraphics[scale=0.2]{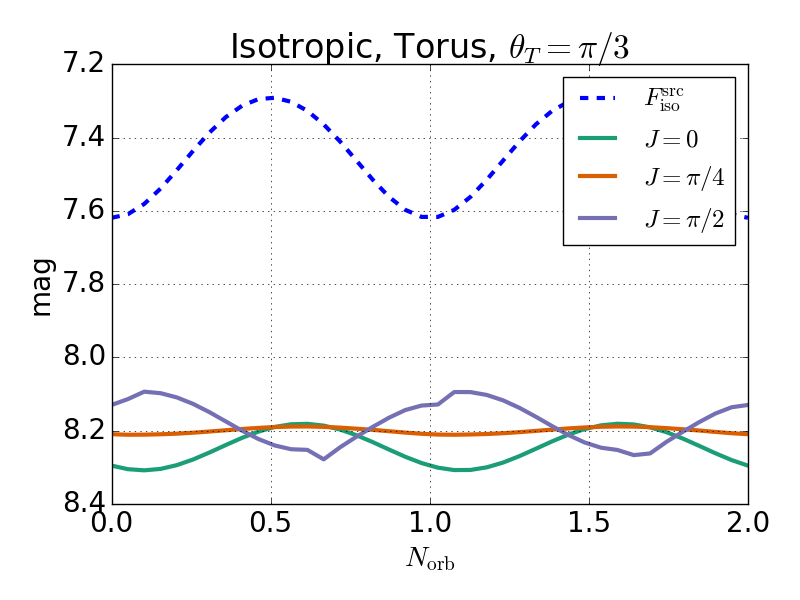} &
\includegraphics[scale=0.2]{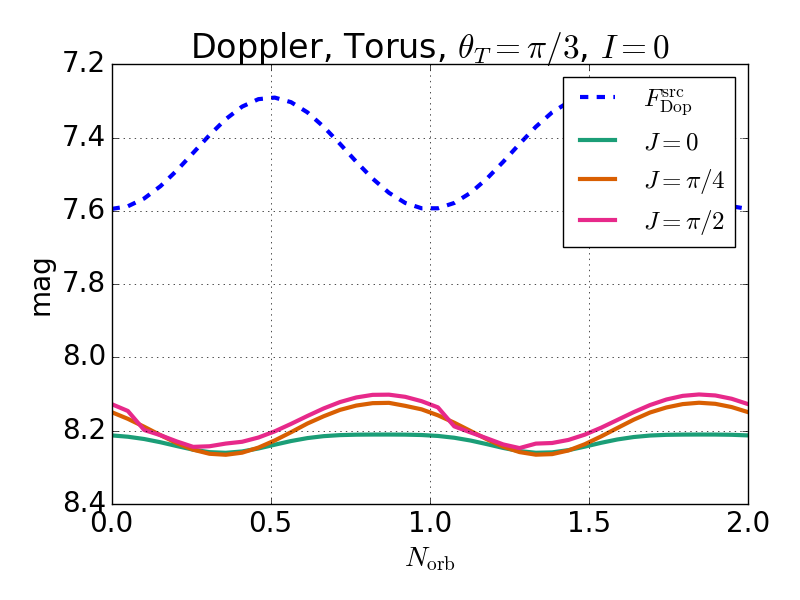}
\end{array}$
\end{center}
\vspace{-15pt}
\caption{The same as Figure \ref{Fig:SphDopvISO} but for the torus
  model.  Here the size of the torus is $R_d = 0.6 R_0$, and each panel
  shows IR echoes for different torus inclinations ($J$), and for torus
  opening angles $\theta_T = \pi/4$ (top) and $\theta_T = \pi/3$
  (bottom). The left (right) panels assume an isotropically variable
  (Doppler-boosted) central source. The (arbitrary) magnitude scale is 
  the same as that used for the spherical dust shell light curves 
  (Figs.~\ref{Fig:SphDopvISO}~and~\ref{Fig:Sph_VarI}).}
\label{Fig:ShTor_DopvISO}
\end{figure}

In Figure \ref{Fig:ShTor_DopvISO} we show the IR echoes (solid lines) and
optical/UV light-curves (dashed lines) for various torus opening and
inclination angles, choosing a fixed value of $t_d/P=0.6$. We recover the
amplitudes shown in Figure \ref{Fig:AIRoAUV_tr} and \ref{Fig:AIRoAUVDop_tr}
and observe the expected dimming of the IR light curves for larger torus
opening angles. In both the isotropic and Doppler cases, the phase lag of the
IR echo is independent of the dust geometry parameters $J$ and $\theta_T$. In
the isotropic cases (left panels in Fig.~\ref{Fig:ShTor_DopvISO}) the 
half-cycle phase shift (\S \ref{S:Simplestcase}) between the $J=\pi/2$ and $J=0$
curves is consistent with the corresponding signs of $A_{\rm{IR}}/A$ in Figure
\ref{Fig:AIRoAUV_tr}: for $J=\pi/2$, $A_{\rm{IR}}/A <0$ for the chosen
$\theta_T$, while for $J=0$, $A_{\rm{IR}}/A > 0$.

\begin{figure}
\begin{center}
\includegraphics[scale=0.4]{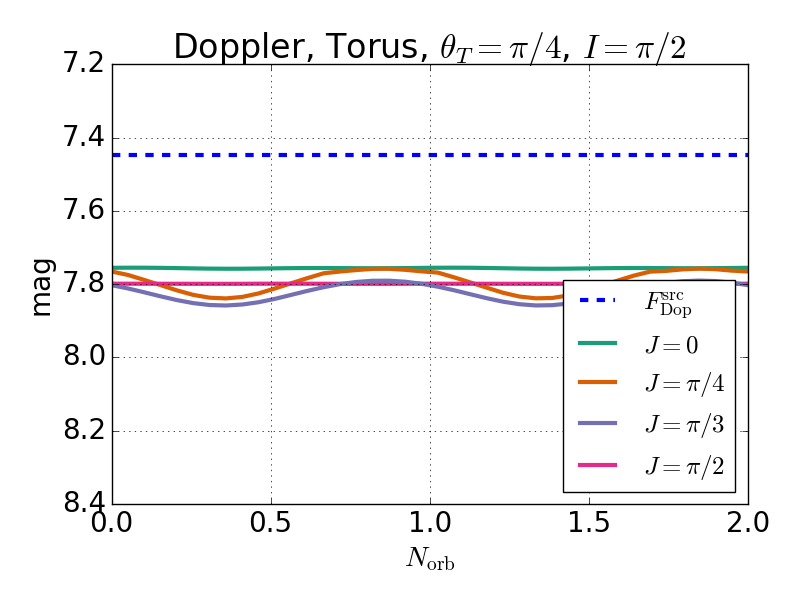} 
\end{center}
\vspace{-15pt}
\caption{The same as Figure~\ref{Fig:ShTor_DopvISO} but for a face-on
  binary.  For all but the most extreme torus inclinations ($J=0$ and
  $\pi/2$), significant IR variability persists even though no
  optical/UV variability is observed.}
\label{Fig:ShTor_VarI}
\end{figure}

Finally, Figure \ref{Fig:ShTor_VarI} explores the effects of binary
inclination in the Doppler torus model. With the freedom to orient the binary
plane relative to a non-spherically symmetric ($\theta_T \neq 0$) dust
structure through parameters $I$ and $J$, the possibility of generating IR
variability with no observed optical/UV variability arises. Figure
\ref{Fig:ShTor_VarI} demonstrates that when the binary is face on, there is no
observed optical/UV variability (\textit{e.g.} Eq.~\ref{Eq:vlos}), but the IR
variability persists.

For the case of a Doppler-boosted source, surrounded by a spherical dust
shell, a face-on binary generates no IR variability because there
is no time-changing emission between the front and back hemispheres of the
dust shell, the integrated dust emission is constant.  This back-to-front
symmetry is broken in the case of the torus, as long as the dust is not
mirror-symmetric around a plane perpendicular to the line of sight and
containing the source, \textit{i.e.}, $\theta_T \neq 0$, $J\neq 0$ and $J\neq
\pi/2$.

\section{Application to MBHB candidate PG~1302-102}
\label{S:PG1302}

We now apply our models to the periodically variable quasar and MBHB candidate
PG~1302-102. We use data published by J15 from the W1 ($3.4 \mu$m) and W2
($4.6 \mu$m) bands of the WISE instrument. Measurements in the WISE W3 ($12
\mu$m) and W4 ($22 \mu$m) bands only exist at two epochs so we do not use
these to fit IR light curves (see further discussion below), however, these
bands can still be used to constrain the spectral energy distribution (SED).
Data for W1 and W2 exist at six-month intervals from 2010 to July 2015 with a
three-year gap during the hibernation of the instrument between 2011 and 2014. As in J15, we
supplement the WISE W1 and W2 band data with an earlier data point from the
AKARI mission \citep[][see J15 for details]{AKARI:2007}. Each epoch of WISE
measurements consists of $\sim 10$ data points taken within a few days of each
other. To fit models to the WISE data, we use a single data point at each
epoch with mean and variance given by the average mean and variance of each
data point at that epoch (see Figure \ref{Fig:SinFit}).

\subsection{Strategy}
Using our models, we aim to determine whether the IR data from PG~1302-102 is
consistent with reverberation from a periodic optical/UV source, and whether
such a source is varying isotropically, or via the Doppler-boost mechanism. We
also aim to constrain the parameters of a putative dusty torus.

Given the optical/UV period $P$ of the source, the primary observables
are the relative brightness, variability amplitude, and phase of the
IR light curves. We have shown that the amplitude ratio
$A_{\rm{IR}}/A$ is dependent on the timescale ratio $t_d/P$ and on the
dust torus opening angle $\theta_T$ and inclination $J$; the IR phase
lag $\Phi$ is dependent on the ratio $t_d/P$. For a Doppler-boosted
source, the IR amplitude also depends on binary inclination $I$ and
orbital velocity (given by the binary mass, mass ratio, and orbital period).

We measure model-independent values of $P$, $\Phi$ and $A_{\rm{IR}}/A$
directly from the optical and WISE light curves. A blackbody spectral fit to
the WISE data yields a measurement of the total IR luminosity and also the
fraction of the optical/UV luminosity echoed in the IR. Together these
determine the size $R_d$ and covering fraction $\cos{\theta_T}$ of the dust
torus, yielding $\theta_T$ and $t_d/P$. The phase lag $\Phi$ provides a
second, independent measurement of $t_d/P$ through the dust echo models.
Because of the sinusoidal nature of the light curve, the values of $t_d/P$
measured from phase lags can only be determined up to addition of an integer
multiple of the variability period. The two measurements of $t_d/P$ determine
whether a consistent dust-echo model is viable and, if they agree, they narrow
the possible values of $t_d/P$. Hence we may use the inferred values of
$t_d/P$ and $\theta_T$ and determine values of the torus inclination $J$ that
yield the observationally determined $A_{\rm{IR}}/A$ (the J dependence of
$A_{\rm{IR}}/A$ is discussed in \S \ref{S:IsoTorAnyl} and \S
\ref{S:DopTorAnyl}). In the Doppler case we choose the binary orbital speed
and inclination based on fits to the optical variability data alone
\citep[\textit{e.g.}][]{PG1302Nature:2015b}. A consistent model yields
constraints on the dusty torus opening angle, inclination, and radius.

\subsection{Model-Independent Measurements}
\label{S:Model-Independent Measurements}

To extract the phase lag and relative amplitude of IR light curves, we fit
separate sinusoids to the V-band, W1, and W2 data.\footnote{The UV data shows
periodicity consistent with the optical period, but has much fewer data
points~\citep{PG1302Nature:2015b}.} For all model fitting we maximize a
$\chi^2$ likelihood using the public Markov Chain Monte Carlo code
\textsc{emcee} \citep{DFM:2013}. Details can be found in Appendix
\ref{A:SinFit}.

\begin{figure}
\begin{center}$
\begin{array}{c}
\hspace{-0.7cm}\includegraphics[scale=0.22]{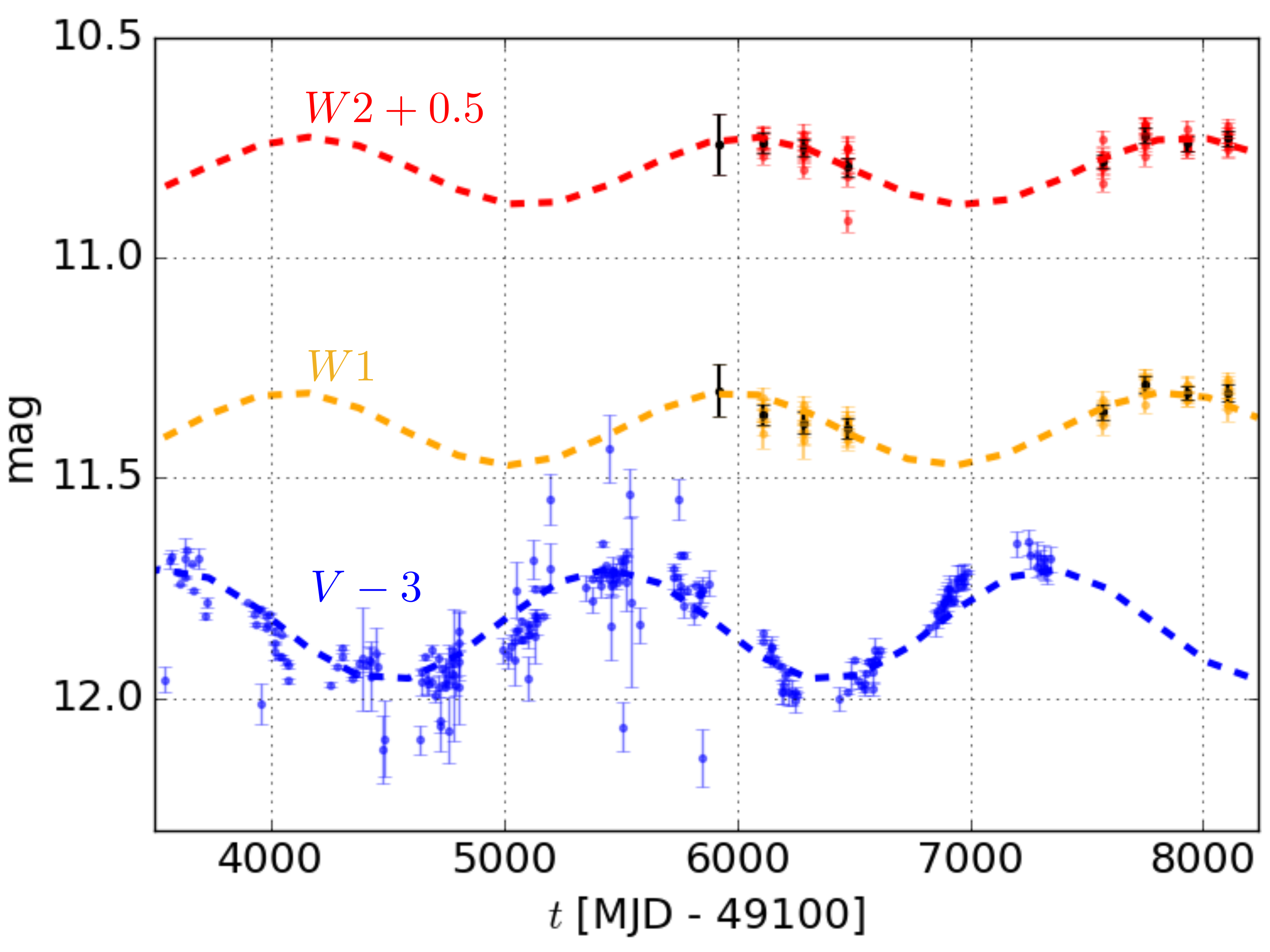}  
\end{array}$
\end{center}
\vspace{-10pt}
\caption{Best-fit sinusoid models to the optical V-band (blue), and
  the IR W1 (orange), and W2 (red) bands. The black points and error
  bars represent the average mean
  and standard deviation of the $\sim$10 IR points at each epoch and
  are used to fit the IR data.  }
\label{Fig:SinFit}
\end{figure}

We proceed by assuming that the IR light curves are caused by reverberation of
the optical/UV and hence have the same period (J15). Fixing the period to the
value observed for PG~1302-102 in the optical, $P=1,884\pm 88$ days, we fit
sinusoids of the form $\mathcal{A} \sin{\Omega\left(t-t_0\right)} +
\mathcal{B}$ to the IR data. The best-fit sinusoids are shown in
Figure~\ref{Fig:SinFit}, along with the V-band, W1, and W2 data. The best-fit
parameters are recorded in Table~\ref{Table:SinParams}.

\begin{table}
\begin{center}
\begin{tabular}{ c | c | c | c }
            		    & V   						   & W1          				  & W2          \\ 
\hline             		   
$\mathcal{A}$ [mag]     & $0.1269^{+0.0004}_{-0.0004}$ & $0.0862^{+0.0158}_{-0.0271}$ &  $0.0813^{+0.0157}_{-0.0260}$    \\&&& \\
$t_0$ [days]            & $298.1^{+0.7}_{-0.7}$    & $715^{+26}_{-32}$      &  $768^{+28}_{-27}$ \\ &&&\\
$\mathcal{B}$ [mag]     & $14.8329^{+0.0003}_{-0.0003}$ & $11.39^{+0.01}_{-0.02}$      &  $10.31^{+0.01}_{-0.02}$  \\&&& \\
$\chi^2_{\rm{red}}$     & 40.5                 & 2.0                  & 0.4      
 \end{tabular}
\caption{
Best-fit sinusoid parameters for model $\mathcal{A}
\sin{\Omega\left(t-t_0\right)} + \mathcal{B}$, with $\Omega = 2\pi/1884$
day$^{-1}$. The reduced $\chi^2$ statistic for the optical fit is consistent
with that found by \citep{PG1302Nature:2015b}, but without the inclusion of
damped random walk noise. The reduced $\chi^2$ statistic is smaller for the W1
and W2 data because we have binned each epoch into one data point as explained
in Appendix \ref{A:SinFit}.
}
\end{center}
\label{Table:SinParams}
\end{table}

The best-fit sinusoid amplitudes in Table \ref{Table:SinParams} give relative 
IR to optical amplitudes
\begin{eqnarray}  \nonumber
\frac{A_{\rm{W1}}}{A_{\rm{V}}}  \approx 0.68^{+0.12}_{-0.21}, \quad
\frac{A_{\rm{W2}}}{A_{\rm{V}}}  \approx 0.64^{+0.12}_{-0.20}. 
\end{eqnarray}

To compare to our analytic solutions we require the quantity $A_{\rm{IR}}/A$.
The relevant quantity ``$A$'' is the variation amplitude of the flux
in the frequency range that dominates the dust heating. In the Doppler case we
obtain $A$ from knowledge that the dust sees a modulation proportional to the
Doppler factor to the fourth power (See Eq. \ref{Eq:DopAll}), so that the 
total optical/UV amplitude is related to $A_V$ by $A = 4/(3-\alpha_V) A_V$.
Using $\alpha_V = 1.1$ \citep{PG1302Nature:2015b}, we find that $A=2.1A_V$.
Then averaging the W1 and W2 values we find
\begin{equation}
\frac{A^{\rm{Dop}}_{\rm{IR}}}{A} \approx  0.31^{+0.07}_{-0.11}. \nonumber
\end{equation}

In the isotropic case we have incomplete knowledge of the frequency
dependence of source variability. The values of $A$ in the V-band ($\approx
13\%$) and the FUV ($\approx 35\%$) are known to differ significantly, but we
know that the slope of the $\lambda F_{\lambda}$ spectrum of PG~1302-102 is
increasing in the FUV, levels off in the NUV, and decreases in the optical
\citep{PG1302Nature:2015b}. Hence, the peak optical/UV power is likely emitted
somewhere in-between the UV and optical values. To be conservative, we
assume that the FUV and optical values bracket the actual effective $A$ by
taking the mean and variance of the four values $A_{W1}/A_V$, $A_{W2}/A_V$,
$A_{W1}/A_{FUV}$, and $A_{W2}/A_{FUV}$,
\begin{equation}
\frac{A^{\rm{Iso}}_{\rm{IR}}}{A} \approx 0.46^{+0.2}_{-0.2}.  \nonumber
\end{equation}

The best-fit phases relative to the V band are
\begin{eqnarray} 
\Phi_{W1} &\approx& 0.284^{+0.017}_{-0.022} + m \nonumber \\
\Phi_{W2} &\approx& 0.319^{+0.019}_{-0.018} + m \quad m=0,1,2...\nonumber \ .
\end{eqnarray}
The phase lags measured in J15 are
\begin{eqnarray} 
\Phi_{W1} &\approx& 0.18 \pm 0.08 + m \nonumber \\
\Phi_{W2} &\approx& 0.28 \pm 0.08 + m \quad m=0,1,2... \nonumber \ .
\end{eqnarray}  
Our values of the phase lag are consistent with J15's, though
the mean of our W1 phase lag extends out of the range measured in
J15. This discrepancy in the phase lags measurements
likely emerges from the two different methods used to determine them: J15
cross-correlate the V-band and IR light curves, while we fit sinusoids to
each. Our smaller error bars likely derive from imposition of a sinusoid model
inherent to our method for estimating $\Phi$. To be conservative, we use
smallest and largest values of the phase lag allowed by the $1 \sigma$ errors
from both measurements to derive a range of values for $t_d/P$ from Eq.
(\ref{Eq:ISOLags}) and (\ref{Eq:DOPLags}).

\subsection{Constraints from Spectral Information Only}

To place further constraints on $t_d/P$, and to constrain the covering
fraction of the dusty torus $\cos{\theta_T}$, we estimate the area and
temperature of the emitting dust. These constraints rely only on the observed
time-averaged SED, and not on the details of the time-dependent dust echoes.
Throughout we assume, as is observed, that PG~1302-102 is an unobscured quasar
so that the inferred optical/UV luminosity can be associated with that which
heats the dust.

We fit blackbody spectra to the time-averaged fluxes (see Appendix
\ref{A:BBfit} for details) by varying the dust temperature $T_d$, and the
total emitting area $4 \pi R^2_d \cos{\theta_T}$ of the dust.\footnote{We have
assumed $Q_{\nu}= 1$, but the result does not greatly differ if we fit for the
efficiency parameters $\nu_0$ and $k$ (see Table \ref{Table:params}).}  We
impose the thermal-equilibrium constraint $R_d = \sqrt{L/(16 \pi \sigma
T^4_d)}$, where we estimate the bolometric luminosity as $L = \rm{BC} \times
L_V$ from the V-band luminosity and from the range of bolometric corrections
determined empirically for quasars with similar brightness, $6 \leq \rm{BC}
\leq 12$ \citep{RichardsQBCs:2006}. Then, given the best-fit value for the
total area and the range of possible BC's, we can solve for the range of
possible covering fractions $\cos{\theta_T}$ and dust radii. Conceptually,
given a best-fit temperature, the range of allowed BC's sets the range of
possible dust sphere radii, then the covering fraction is constrained to match
the total observed IR luminosity.

An important source of ambiguity is that we do not know what fraction
of detected IR flux in each band is attributable to reverberation of
the central source luminosity -- as opposed to spatially unresolved
emission from unrelated off-nuclear dust heated by star-formation, or
any other source of IR emission unrelated to the torus. We here take
two approaches to fitting the IR SED:
\begin{enumerate}
\item We attempt to fit all four WISE bands, and find that a blackbody
  spectrum can be fit to the W1, W2, and W3 bands, but not W4. We are
  forced to assume that contributions from W4 are primarily \textit{not} due to
  reverberation of the central source continuum.
\item We fit only the W1 and W2 band fluxes, because these are the
  only two bands for which measurements of variability exist,
  confirming the same periodicity as in the optical/UV.  These two
  bands should therefore be associated with reverberation of the
  central source, but without data at more epochs, this remains
  unclear for the W3 band.
\end{enumerate}

For case (i), using the W1, W2, and W3 fluxes, we find a spectrum with 
best-fit parameters  $T_d = 577 \pm 3$K, $\cos{\theta_T} = 0.08^{+0.04}_{-0.03}$,
and $R_d = 3.9 \pm 1.0$ pc, corresponding to $t_d/P = 3.1 \pm 0.8$.

For case (ii), using only W1 and W2 fluxes, we find a spectrum with best-fit
parameters  $T_d = 881 \pm 14$K, $\cos{\theta_T} = 0.05^{+0.02}_{-0.02}$, and
$R_d = 1.9 \pm 0.4$ pc, corresponding to $t_d/P = 1.5 \pm 0.3$.

Case (i) yields a solution for dust that is colder, further out, and
has a larger covering fraction than in case (ii). This is because
there is more flux in the lower-wavelength W3 band, pushing
case (1) to a lower temperature but a larger covering fraction (see
Figure \ref{Fig:BBFit}).

A clear way to distinguish between the above two cases, and 
to remove the ambiguity in the best-fit values for the reverberated IR
luminosity, is to continue monitoring the longer-wavelength emission in
the W3 (and W4) bands for periodicity.

Table \ref{Table:IRmeasures} summarizes our measured values for the IR 
PG~1302-102 quantities. We now use them to constrain dust and source
properties of PG~1302-102.

\begin{table*}
\begin{center}
\begin{tabular}{ c | c | c | c | c}
\hline
\hline
\multicolumn{5}{|c|}{\it I Model Independent Measures }\\
\hline
\hline
$\frac{A_{\rm{W1}}}{A_V}$  &  $\frac{A_{\rm{W2}}}{A_V}$ &  $\Phi_{\rm{W1}}$ [cycles]  & $\Phi_{\rm{W2}}$ [cycles] & $P$ [days]\\
\hline
$0.68^{+0.12}_{-0.21}$ & $0.64^{+0.12}_{-0.20}$ & $0.10 \rightarrow 0.30$ & $0.20 \rightarrow 0.36$ & $1877^{+22}_{-3}$ \\
                       &                        & \multicolumn{2}{|c|}{ $\{\pm$ Half Integer $\}$} &  \\
\hline
\hline
\multicolumn{5}{|c|}{\it II Spectral Fit Measures }\\
\hline
\hline
            & $T_d$  [K]        &   $\cos{\theta_T}$     &   B.C.       &  $R_d$ [pc] \\
            \hline
case (i)    &  $577 \pm 3$  & $0.08^{+0.04}_{-0.03}$ &  $9\pm3$  & $3.9 \pm 1.0$ ($t_d/P =3.1 \pm 0.8$) \\
case (ii)   &  $881 \pm 14$     & $0.05^{+0.02}_{-0.02}$ &  $9\pm3$ & $1.9 \pm 0.4$ ($t_d/P =1.5 \pm 0.3$)  \\
\hline
\hline
\multicolumn{5}{|c|}{\it III Dust-Echo Modeling Measures }\\
\hline
\hline
        &  $\frac{A_{\rm{IR}}}{A}$  & $t_d/P$          &   $J$ [rad]           &  $I$ [rad]  \\
\hline
Iso, case (i)    &  $0.46 \pm 0.2$ &  $\{2.3\rightarrow2.36$ &  $J \gsim 1.0$  &   Unconstrained  \\
&&                                    $2.6\rightarrow2.86$   && \\
&&                                    $3.1\rightarrow3.36$   && \\
&&                                    $3.6\rightarrow3.86\}$   && \\
Iso, case (ii)    &&                  $\{1.2\rightarrow1.36$ &   $1.1\lsim J \lsim 1.5 \cup J \lsim 0.6$ &   Unconstrained  \\
&&                                    $1.6\rightarrow1.8\}$   &&  \\
\hline 
Dop, case (i)    &  $0.31^{+0.07}_{-0.11}$  & $\{2.35\rightarrow2.61$  &  Not allowed for $I=0$ &   $I<0.7$  \\ 
&&                                           $2.85\rightarrow3.11$     &                 & (Figure 1 \cite{PG1302Nature:2015b})\\
&&                                           $3.35\rightarrow3.61$    && \\
&&                                           $3.85\rightarrow3.9\}$     && \\
Dop, case (ii)    &&                  $\{1.35\rightarrow1.61\}$  & $J \lsim 0.6$  &   
\end{tabular}
\caption{
Measured quantities used to determine the nature of the IR light curves of PG~1302-102.
The table is divided into three sections.
Section I displays the quantities measured directly from the light curves with
no model assumptions: the ratios of IR to V-band amplitudes, phase lags, and
period.
Section II displays quantities found from fitting a blackbody
spectrum to WISE band fluxes. Case (i) utilizes W1, W2, and W3 bands,
case(ii) uses only W1 and W2 bands.
Section III displays quantities found from incorporating the quantities
measured in the first two sections of the table into our dust-echo models.
Different results are found depending on the source model: an isotropically
varying source (Iso) or a Doppler-boosted source (Dop), and the data used in
the spectral fit: case (i) and (ii) above. Note that $t_d/P$ is constrained
both by the $\Phi$ measurements in the first section of the table and also by
$R_d$ in the second section. Because the $\Phi$ measurements yield $t_d/P$
values modulo half of a variability cycle (Eqs. \ref{Eq:DOPLags} and
\ref{Eq:ISOLags}), the combined constraint on $t_d/P$ can consist of multiple
disconnected regions. The $A_{\rm{IR}}/A$ value in section III combines both
relative amplitude measurements in the first section of the Table and the
source type which sets $A$.
}
\end{center}
\label{Table:IRmeasures}
\end{table*}

\begin{figure*}
\begin{center}$
\begin{array}{c}
\includegraphics[scale=0.33]{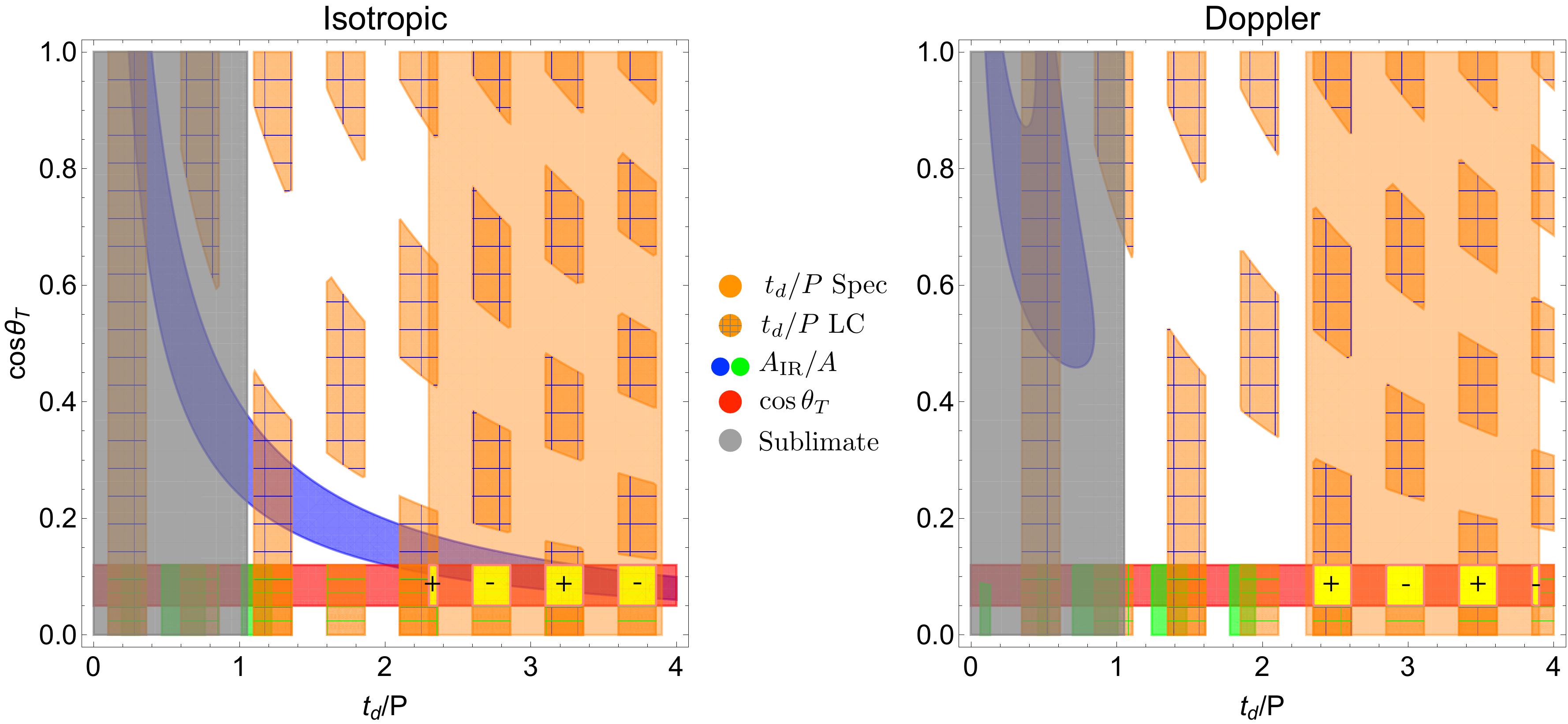} 
 \\ 
\hspace{7pt}
\includegraphics[scale=0.33]{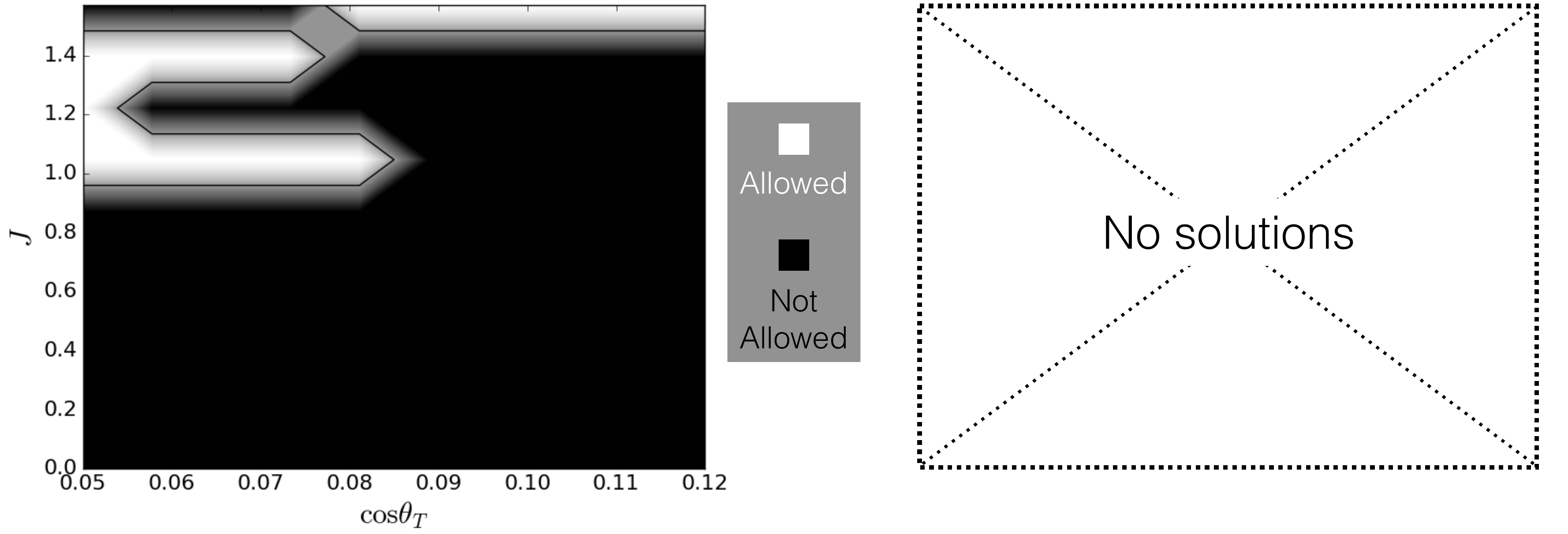}
\end{array}$
\end{center}
\vspace{-10pt}
\caption{
Regions of allowed parameter space for dust torus models surrounding a MBHB in
PG~1302-102. The left (right) panels are for isotropically variable (Doppler
boosted) sources. \textit{Top~row:} Regions allowed in ($\cos{\theta_T}$,
$t_d/P$) parameter space. The grey-shaded vertical bands on the left are
excluded because the temperature is above the sublimation value for graphites
(but see text). Blue (green) regions show example dust models where the IR
echo variability amplitude $A_{\rm IR}$ from a face-on torus (edge-on thin
ring) is large enough to match the observed value. The horizontal red bands
show the range of the dust covering fraction $\cos{\theta_T}$ allowed by IR
SED fitting. The hashed orange regions show constraints on $t_d/P$ from the
phase lags predicted for a face-on torus (with blue hash) or edge-on ring
(with green hash). Because the green regions denote solutions for which
$\cos{\theta_T} \rightarrow 0$, we only plot them for small $\cos{\theta_T}$.
The solid orange vertical band is from the independent measurement of $t_d/P$
from the IR~SED~fits. Yellow rectangles highlight regions allowed by all of
the $\cos{\theta_T}$ and $t_d/P$ constraints. The yellow regions do not
include $A_{\rm{IR}}/A$ constraints but are marked with a ``+'' or ``-'',
indicating the required sign of $A_{\rm{IR}}/A$ in this region (determined by
Eqs. \ref{Eq:ISOLags} and \ref{Eq:DOPLags}). \textit{Bottom row:} Focusing
only on the allowed yellow regions of the top panels, the bottom panels show
the torus inclination angles $J$ for which $A_{\rm{IR}}/A$ matches the
measured values, and hence, where all five constraints are met (white
regions). The bottom right panel would be entirely black so we do 
we simply state that no solutions exist in this case.
}

\label{Fig:PGConts_W1W2W3}
\end{figure*}

\subsection{Interpretation with Dust-Echo Modeling}

We now place constraints on dust torus models, based on the
predicted properties of their dust echoes.

In Figure~\ref{Fig:PGConts_W1W2W3}, we show 
regions allowed by various constraints in 2D parameter planes:
in ($\cos{\theta_T}$, $t_d/P$) in the top row and in the
($J$, $\cos{\theta_T}$) plane in the bottom row.
The left panels are for isotropically varying sources and the right
panels are for Doppler-boosted sources.  

\begin{figure*}
\begin{center}$
\begin{array}{c c}
\includegraphics[scale=0.33]{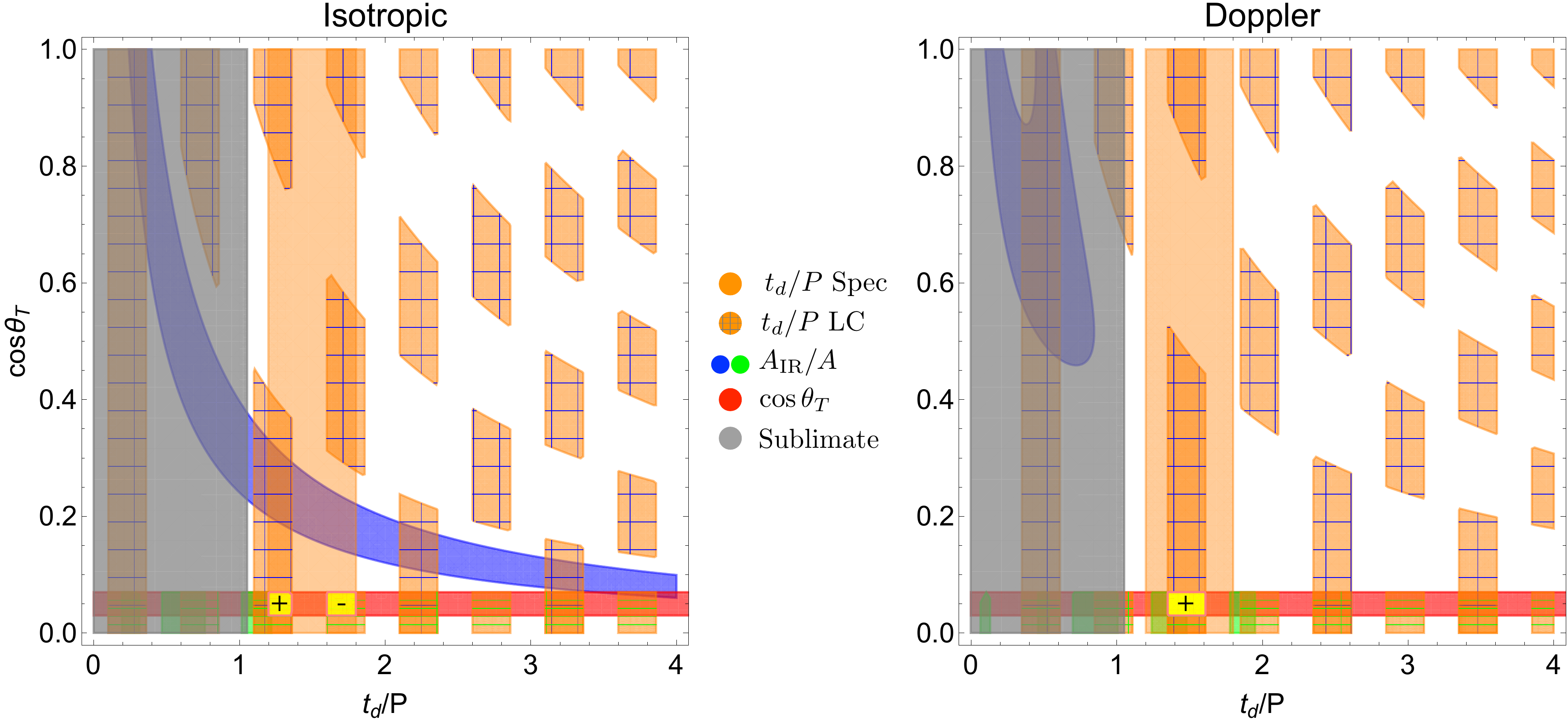} \\ 
\hspace{7pt}
\includegraphics[scale=0.33]{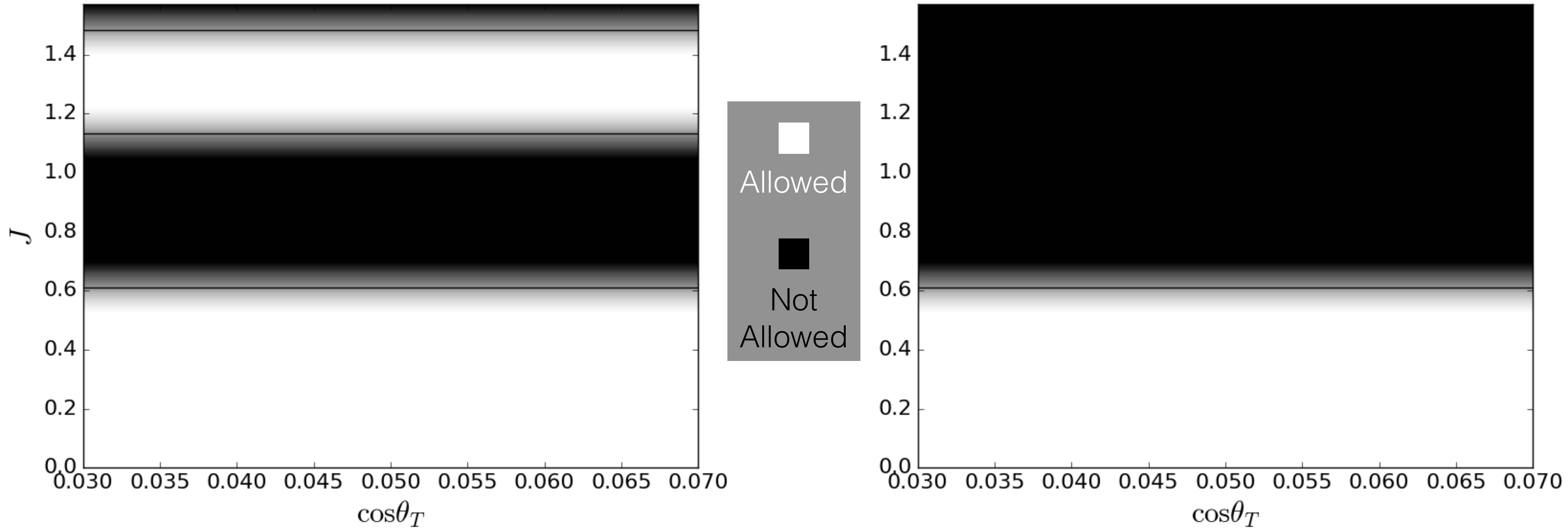}
\end{array}$
\end{center}
\vspace{-10pt}
\caption{The same as Figure \ref{Fig:PGConts_W1W2W3} but using
  quantities derived from SED fitting only the time-averaged WISE W1
  and W2 band fluxes. These are the only two bands with sufficient
  time-domain data and which show evidence for periodic variability.}
\label{Fig:PGConts_W1W2}
\end{figure*}

In the top row, we show five distinct constraints:
\begin{enumerate}
\item The regions where the values of $A_{\rm{IR}}/A$
predicted by our models fall within the measured values (Table
\ref{Table:IRmeasures}) are shaded blue and green, for two different
fixed values of the torus inclination for which the solutions are
analytic. The blue regions are for
a face-on torus ($J=\pi/2$; Eq.~\ref{Eq:LIR_thT})
and the green regions are for an edge-on dust ring ($J=0$ and
$\theta_T\approx \pi/2$; Eqs.~\ref{Eq:ISO_Ring}
and~\ref{Eq:DOP_Ring}).  
There are multiple disjoint allowed
regions, because of the oscillatory patterns in $A_{\rm{IR}}/A$ as a
function of $t_d/P$.

\item The horizontal red bands show the allowed range of the covering fraction
$\cos{\theta_T}$ inferred from the IR SED modeling alone. This
constraint is independent of $t_d/P$.

\item The hashed, orange regions show the values of the ratio $t_d/P$ allowed by
fitting the observed phase lags using the dust echo models
(Eqs.~\ref{Eq:ISOLags} and \ref{Eq:DOPLags}).  The disjointed nature
of these orange regions is from the ambiguity in adding an integer
number of periods to the phase lag measurements, and also from the
half-cycle phase change occurring when $A_{\rm{IR}}/A$ changes
sign. Because the sign of $A_{\rm{IR}}/A$ depends on $\cos{\theta_T}$
and $J$, we hash the orange regions with the color corresponding to
the value of $J$ for which they are calculated: blue ($J=\pi/2$) and
green ($J=0$). 

\item The vertical solid orange band spanning the entire y-axis is
from the independent measurement of $t_d=R_d/c$ from the blackbody fits 
to the average WISE-band fluxes (see Table \ref{Table:IRmeasures}),
combined with $P$ measured directly from the sinusoidal fits to the
light-curves.

\item By assuming that the source emits at $\epsilon$ of the Eddington
  luminosity, we estimate the sublimation radius of dust as a function
  of black hole mass (using Eq.~2 of \citealt{MorNetzer:2012}). Assuming
  that dust tori cannot exist inside this radius, we obtain the lower limit
\begin{equation}
\frac{t_d}{P} \gsim 1.0 \left(\frac{\epsilon}{0.5} \right)^{1/2} \left(\frac{P}{4 \rm{yr} } \right)^{-1} \left(\frac{M}{10^9 \Msun } \right)^{1/2}.
\end{equation}
The grey shaded regions to the left are excluded by this constraint (in Figure
\ref{Fig:PGConts_W1W2W3} we use the PG 1302 bolometric luminosity to determine
the sublimation region).

\end{enumerate}

The top panels shows that there exist consistent models for periodically
reverberated emission from PG~1302-102 which satisfy all five constraints.
Graphically these are realised where the red, solid orange, blue-hashed
orange, and blue ($J=\pi/2$) \textit{or} red, solid orange, green-hashed
orange and green ($J=0$) regions overlap, and avoid the grey regions. However,
these overlaps do not express the full space of solutions, they only take into
account specific dust torus models for which we have analytic solutions. To
show the full space of allowed solutions, we next numerically calculate
$A_{\rm{IR}}/A$ over the range of joint constraints on $t_d/P$ and
$\cos{\theta_T}$ and for the full range of torus inclinations angles $J$.

The regions of allowed values of $t_d/P$ and $\cos{\theta_T}$, for which we
now calculate $A_{\rm{IR}}/A$ as a function of $J$, are highlighted in yellow
in the top panels of Figure \ref{Fig:PGConts_W1W2W3}. The bottom panels of
Figure \ref{Fig:PGConts_W1W2W3} calculate $A_{\rm{IR}}/A$ in only these yellow
regions, for a grid of torus inclinations spanning from $0$ to $\pi/2$ and the
allowed range of $\cos{\theta_T}$. For each pair of $J$ and $\cos{\theta_T}$,
we ask if any value of $t_d/P$ in the allowed range generates $A_{\rm{IR}}/A$
in the measured range; if so, we assign the region in the bottom panels of
Figure \ref{Fig:PGConts_W1W2W3} the color white, if not that region is colored
black. We then draw a black contour excluding the black regions. In this way
we discern the limits on the torus inclination in each of our models.
Note the we also take into account that each yellow region, in
progressing $t_d/P$, is allowed only for a specific sign of $A_{\rm{IR}}/A$ (see
Eqs. \ref{Eq:ISOLags} and \ref{Eq:DOPLags}), and this sign alternates between
yellow regions (see ``+'' and ``-'' labels in the top rows of Figures
\ref{Fig:PGConts_W1W2W3} and \ref{Fig:PGConts_W1W2}).

For computational simplicity, to calculate $A_{\rm{IR}}/A$ in the
bottom panels of Figure \ref{Fig:PGConts_W1W2W3}, we allow $Q_{\nu} =1$
everywhere and compute the IR variability amplitude when integrating over all
frequencies. As Figure \ref{Fig:Qvon} shows, this generates a small
overestimate for the relative IR amplitude and results in a conservative
estimate of the ruled out dust parameters (overestimate of the size of white
regions in Figures \ref{Fig:PGConts_W1W2W3}). Practically, we use a
$10\times10$ grid of $J$ and $\cos{\theta_T}$, sampling each yellow region
uniformly with five values of $t_d/P$. With these approximations in mind, we
use only the general trends in the bottom panels of Figure
\ref{Fig:PGConts_W1W2W3} to estimate the regions of excluded torus inclination
angle.

Figure \ref{Fig:PGConts_W1W2} illustrates the same restrictions on parameter
space as Figure \ref{Fig:PGConts_W1W2W3}, but for the constraints found by
including only the W1 and W2 band average fluxes in the SED fitting (section
II, case (ii) in Table \ref{Table:IRmeasures}).


Figures \ref{Fig:PGConts_W1W2W3} and \ref{Fig:PGConts_W1W2} reveal that the
short-wavelength ($\lsim 20 \mu$m) IR emission from PG~1302-102 is consistent
with reverberation by a periodically variable source. However, only the case
which includes W1 and W2 bands is consistent with all constraints assuming an
isotropically varying or a Doppler-boosted central source. Case (i) which also
includes SED information from the W3 band is consistent only with an
isotropically varying source, for a small range of nearly face-on dust tori.
The allowed parameter values depend on whether the WISE W3 band is excluded or
included in the SED fitting, along with the W1 and W2 bands. Common to both
cases is the restriction to a small dust covering fraction (requiring the
dusty torus to resemble a thin disc). This constraint arises from fitting the
time-averaged WISE SEDs, independently of the time-dependent dust echo
modeling.

When we include the W1, W2, and W3 bands in the SED fit (section II,
case (i) of Table~\ref{Table:IRmeasures} and Figure \ref{Fig:PGConts_W1W2W3})
we find a dust covering fraction, or equivalently a disc aspect ratio, of
$0.05 \lsim \cos{\theta_T} \lsim 0.12$, and inner radii of $\sim 2.9
\rightarrow 4.9$ pc, but for different disjoint values indicated by the ranges
of $t_d/P$ listed in \S III of Table \ref{Table:IRmeasures} for isotropic and
Doppler-boost models. The bottom panels of Figure \ref{Fig:PGConts_W1W2W3}
show us that, in the isotropic case, the inclination of the dusty disc must be
$J\gsim 0.8$ radians ($57^{\circ}$) away from edge on, and may be face on. In
the Doppler case, there are no allowed dust torus solutions within our model.

That there are consistent solutions for periodically reverberated
emission from an isotropic source, but not a Doppler-boosted source in case
(i) may simply be a manifestation of the greater uncertainty on the value of
$A_{\rm{IR}}/A$ for the isotropic source (see \S~\ref{S:Model-Independent
Measurements}). A better determination of variability amplitudes across the
spectrum of PG 1302 will help to resolve this issue.

Including only the bands for which there is evidence for periodic variability,
the W1 and W2 WISE bands (section II, case (ii) of Table
\ref{Table:IRmeasures} and Figure \ref{Fig:PGConts_W1W2}) we find dusty discs
with aspect ratio $0.03 \lsim \cos{\theta_T} \lsim 0.07$ and and inner radii
of $\sim 1.5 \rightarrow 2.3$ pc, but for different disjoint values indicated
by the ranges of $t_d/P$ listed in \S III of Table \ref{Table:IRmeasures}. The
bottom panels of Figure \ref{Fig:PGConts_W1W2} show us that, in the isotropic
case, the inclination of the dusty disc is largely unconstrained but cannot be
face on, $J \lsim 0.6$ ($34^{\circ}$) or $1.1 \lsim J \lsim 1.5$ radians
($63^{\circ} \rightarrow 86^{\circ}$), this is also clear in the top left
panel of Figure \ref{Fig:PGConts_W1W2}, where the analytic face-on torus
solutions, represented by the blue region, do not intersect the yellow
highlighted regions. The situation is similar for the Doppler case, the dust
inclination is limited to $J \lsim 0.6$ radians ($34^{\circ}$) from edge on.

In both case (i) and case (ii), the Doppler solutions are restricted from
having face-on dust inclination angles because a thin, face-on dust disc can
not generate large enough reverberated IR amplitudes. This follows because the
dust is constrained to be in a thin disc (torus with a large opening angle).
When such a disc is oriented nearly face on, the time delay along the line of
sight dust structure becomes small and, as discussed in \S\ref{S:DopTorAnyl},
this results in the relative amplitude of IR variability to also become small
for a dust distribution that is symmetric about the origin.

For the isotropically pulsating source, we find that case (i) rules out dust
discs that approach edge-on inclinations, while case (ii) rules out those
approaching face on. This difference is caused mainly by the different
required locations of the dust radius in either case.

In (isotropic) case (i), a more-edge-on ring cannot generate a large enough
relative IR amplitude at the required, large values of $t_d/P$. A face-on dust
torus, however, can generate large relative IR amplitudes for arbitrarily
large $t_d/P$, as long as the dust torus is thin enough, ($\cos{\theta_T}$ is
small; see left panel of Figure \ref{Fig:AIRoAUV_tr}). In (isotropic) case
(i), the relative IR amplitude even becomes too large at small
$\cos{\theta_T}$ as evidenced by a change from allowed to not-allowed face-on
dust tori with decreasing $\cos{\theta_T}$ (see the top of the bottom left
panel of Figure \ref{Fig:PGConts_W1W2W3}).

In (isotropic) case (ii), the smaller radius of the dust torus, and hence, smaller
required values of $t_d/P$, cause relative IR amplitudes to be
too large in the case of a thin, face-on dust torus. At the same time, the
relative IR amplitudes generated by an edge-on ring are large enough to meet
the measured values of $A_{\rm{IR}}/A$ at these smaller values of $t_d/P$.

While the case (ii) spectral fit is more restrictive on the size
of the inner dust region $R_d$, it is less restrictive in dust inclination
angle $J$. This is because it is easier to match the required values of
$A_{\rm{IR}}/A$ for the smaller values of $t_d/P$ in case (ii).

We have so far ignored the relative motion of the binary with respect to the
dusty torus. This motion can introduce additional time-dependence in the
light-travel times, with significance on the order of the ratio of binary
separation to the size of the IR reprocessing region (the dust). This ratio is
\begin{equation}
\frac{a}{R_d} \simeq 0.007\ \epsilon^{-1/2} \left(\frac{M}{10^9 \Msun}\right)^{-1/6}  
\left(\frac{P}{4 \rm{yr}}\right)^{2/3}  \left(\frac{T}{1800 \rm{K}}\right)^{2.8},
\label{Eq:aORd}
\end{equation}
where we have again estimated the dust radius by the sublimation radius (Eq.~2
of \citealt{MorNetzer:2012}) with $\epsilon$ the Eddington fraction. This
tells us that the impact of binary orbital motion on dust reverberation models
is most important for the lowest mass binaries with the longest periods, and
contributes at most at the percent level for the fiducial values taken here
for a PG~1302-102-like binary. Because the dependence on mass is weak, and 5
years is a present upper limit on observed binary periods discovered in EM
time-domain surveys \citep[\textit{e.g.}][]{Graham+2015b}, it is safe to
assume that binary orbital motion is a $\lsim 2\%$ effect.  It will
nevertheless be important to take this effect into account for longer-period
binaries, with orbits still surrounded by a single dust region, binaries
emitting at a small fraction of Eddington, and also for modeling reverberation
by the closer-in broad-line regions around MBHBs (the subject of future work).

Finally, we reiterate that the constraints for thin dust tori around
PG~1302-102 arise from two independent lines of reasoning. The first is based
only on the time-averaged SED data and the geometry of the torus, and is
independent of the time-dependent dust-echo modeling. The black-body emitting
dust must be heated to the observed temperature and emit at the observed
overall luminosity. We find that this forces the dust to be far enough from
the central source to be cool enough to match the IR SED shape, and to also
have a small enough covering fraction to avoid over-producing the IR
luminosity (See section II of Table \ref{Table:IRmeasures}). The second
requirement is based on the light-curves and the time-dependent dust-echo
models. The large measured IR variability amplitudes, and our finding that
thinner dusty discs generate larger relative IR amplitudes is also consistent
with tori being thin, this can be seen, \textit{e.g.}, from the blue regions
in the top panels of Figures \ref{Fig:PGConts_W1W2W3} and
\ref{Fig:PGConts_W1W2}: in the allowed range of $t_d/P$ values,
$\cos{\theta_T}$ is restricted to, at thickest, $\cos{\theta_T}\lsim0.3$ in
the isotropic case (ii).

\section{Discussion and Conclusions}
\label{S:Discussion}
We summarize our key results and discuss their implications for MBHBs.

\begin{enumerate}
\item{ 

\textbf{The phase lag of IR variability} relative to optical/UV
variability is given by $2 \pi t_d/P$ radians in the isotropic case
and $(1 + 1/4)2 \pi t_d/P$ radians in the Doppler case. There is an
additional half-cycle phase shift for light curves which are
reflection symmetric (sinusoids); whether this phase shift occurs
depends on the values of $t_d/P$ as well as the dust geometry (see
Eq. (\ref{Eq:LIR_thT})). These differences, for periodic sources,
should be considered when relating an IR time lag with the light
travel time across the dust reverberation region.

}

\item{   

\textbf{The amplitude of IR variability} relative to optical/UV is a function
of $t_d/P$, the
inclination and opening angles of the dusty torus, and also the binary
inclination to the line of sight for a Doppler-boosted source (see Figures
\ref{Fig:AIRoAUV_tr}--\ref{Fig:ShTor_VarI}). In the isotropic case, the IR
amplitude falls to zero for $t_d/P \gg 1$, and approaches that of the
optical/UV continuum for $t_d/P \rightarrow 0$.  The Doppler case obtains peak
amplitude at $t_d/P \gsim 0.3$, depending on the torus
properties, and falls to zero for both $t_d/P \rightarrow 0$ and $t_d/P \gg
1$. The isotropic case also exhibits zero relative IR amplitude when the
line-of-sight light crossing time of the dust is equal to an integer multiple
the source variability period while the Doppler case exhibits zero relative IR
amplitude when the dust line-of-sight light crossing time is approximately an
integer plus one quarter cycle of the binary orbital period.

When the dust has the geometry of a face-on or edge-on torus, the IR variability
amplitude in the Doppler-boosted case does not exceed $\sim60\%$ of the
optical/UV amplitude (See Figures \ref{Fig:AIRoAUVDop_tr} and
\ref{Fig:AIRoAUV_Rings}). As we showed in Figure \ref{Fig:ShTor_VarI}, this is
not the case when the torus is misaligned and the binary is nearly face on. We
discuss the implications of this in the next bullet point.

While Figure \ref{Fig:AIRoAUV_Rings} shows that thin, nearly-face-on dust
rings can generate large IR variability at any value of $t_d/P$ (though in a
trade-off for total IR luminosity), we find that for the majority of dust and
binary parameters, the largest IR variability amplitudes occurs for smaller
values of $t_d/P$. Hence we use $t_d/P$ as an indicator for IR variability
amplitude in Figure \ref{Fig:boostParams}. In Figure \ref{Fig:boostParams}, we
overlay contours of $t_d/P$ to illustrates MBHB parameters conducive to
generating significant reverberated IR variability.

We conclude from Figure \ref{Fig:boostParams} that, for isotropic sources,
long-period optical/UV modulations by emission from low-mass binaries have the
potential to generate the largest \textit{relative} IR modulations.  The
situation is different in the Doppler case where $t_d/P \lsim 0.1$ results
in small relative IR variability amplitude. In addition to smaller relative
amplitudes, small $t_d/P$ also implies weak absolute optical/UV variability
amplitudes due to the Doppler boost and hence small absolute IR amplitude as
well. Hence, in the Doppler case, either intermediate
masses and binary periods, or the thin nearly-face-on dust rings discussed
above are favored for detection of reverberated IR variability. }

\item{

\textbf{Orphan-IR variability} can occur for Doppler-boosted sources which are
nearly face on (so we do not see the optical/UV variability), but are
surrounded by a dusty torus that is not symmetric across the plane separating
the front and back of the Doppler-boosted source. Periodic orphan-IR
variability in quasars would be a signature of Doppler-boosted MBHBs at high
binary inclination to the line of sight, or to a variety of variable sources
completely surrounded by dust. These periodic sources would not yet have been
identified in optical searches such as those carried out by
\cite{Graham+2015b} and \cite{Charisi+2016}.

}

\item{ 

\textbf{Periodic IR variability from PG~1302-102} is consistent with reverberation
of a Doppler-boosted source surrounded by a thin disc of dust with inclination
within $\lsim 34^{\circ}$ from the line of sight. Consistent solutions for
periodic reverberation also exist for an isotropically varying source
surrounded by a thin dust disc at any inclination of the disc to the line of
sight. Each scenario requires a dusty disc aspect ratio $\sim 0.1$ and inner
radius between 1 and 5 pc. The isotropic and Doppler-boost cases each require
different inner dust radii in this range, allowing a test to differentiate the
source nature if more complete IR SEDs of reverberated emission and phase lag
measurements can narrow down the true size.

Because of the small inferred angular size of the reverberated-IR emitting
region ($\lsim$  milli-arcsecond), it is not currently possible to image this
region in the IR at the distance of PG~1302-102. Though it would be worth
determining if imaging of regions further out in the dust structure could lend
clues to the dust geometry and orientation at smaller radii.

These scenarios also require that longer wavelength ($\gsim 22 \mu$m
for isotropic source and $\gsim 12 \mu$m for a Doppler-boosted source) IR
emission from PG~1302-102 is not from reverberation of the central MBHB
emission, and so our models predict that IR emission in the WISE W4 and
longer wavelength bands will not exhibit periodic variability at the optical
period. A clear determination of which IR wavelengths exhibit such
periodicity, and are thus due to central source reverberation, will help to
constrain the dust+source model for PG~1302-102. Specifically, determination
that the W1 through W3 band ($\sim 3 \rightarrow 17 \mu$m) emission is primarily due
to reverberation by the central source while W4 band emission ($22 \mu$m) is
not, would add evidence against the Doppler-boost nature of the central
periodic source of optical/UV emission.} The enhanced constraints on dust
parameters from narrowing down the reverberated IR SED can be seen by
comparing the constraints from the two possible SED fits in Table
\ref{Table:IRmeasures}, of which we quote only the weaker, joint constraints
above. Further significant improvements would come from measurements of the
frequency dependent variability amplitude of PG~1302-102. These would allow a
tighter constraint on the relative IR to optical/UV amplitude in the isotropic
case.

\end{enumerate}

\subsection{Caveats and extensions}

\begin{itemize}

\item{ 
We have considered smooth, single-species dust models which are optically thin
to their own emission and optically thick to optical/UV emission, and are at a
single temperature at each location. Future work should consider dust models
which include dust heating by optical/UV over a finite radial extent, IR
absorption by dust, clumpy dust \citep[\textit{e.g}][and references
therein]{Netzer:2015:rev}, and a distribution of dust grain species and sizes,
as well as a finite range of temperatures.
}

\item{
The geometry of the dust torus may not be a simple conical cut of a sphere,
but may be convex in nature so that optical/UV light could be absorbed by the
dust at a range of radii even under the assumption here that the dust is
optically thick to optical/UV \citep[\textit{e.g.}][]{NunezHass:2014}. This
could add a colder IR component with significantly different time lag
structure and should be considered in future work.
}

\item{ 
We have considered a dust distribution which is always centered on the
emitting source, and reflection symmetric about the origin. Deviation from
this symmetry will alter our predictions. For example, the Doppler boost IR
amplitudes will not go exactly to zero for $t_d/P \rightarrow 0$ if dust is
only in front of or only behind the source. 
}

\item{
We have considered the case of isotropically variable emission as a
proxy for variations due to accretion changes and also as a control with which
to compare the Doppler-boosted source models. However, emission from an accretion disc
itself is likely spatially anisotropic, emitting less from the disk edges
\citep[\textit{e.g}][and references
therein]{NamekataUmemura:2016}. Unlike the Doppler-boost case, however, this anisotropy is
constant in time. Addition of this effect would introduce another parameter to
both Doppler-boost and non-Doppler-boost models: the relative inclination of
accretion disc and dust torus.
}

\item{  We assume the binary is on a circular orbit. Some hydrodynamical
models of the binary interaction with a gas disc predict the excitation of
large binary eccentricities
\citep[\textit{e.g.}][]{Cuadra:2009,Roedig:2011:eccevo}. Binary eccentricity
could change the shape and periodicity structure of the optical and hence the
IR light curves predicted here (although we note that the nearly sinusoidal
light curve in the example PG~1302-102 shows no evidence for
eccentricity;~\citealt{PG1302Nature:2015b}). }

\item{
If grains can re-form on a timescale shorter than a binary orbital
time, the sublimation radius will change periodically with the
changing central source flux. The change in dust
sublimation due to the changing observed luminosity variations $\delta L$ is
$2 \delta R_d/R_d =  \delta L/L$.
For typical Doppler luminosity variations, this corresponds to a change in the
inner dust radius of a few to $\sim 10 \%$. In the Doppler case, the largest
variations occur for less massive binaries with shorter periods. In the
isotropic case, only the amplitude of modulation is relevant. In either case,
the dust will emit at a constant (source-frame) temperature, but light travel
time lags will acquire extra time dependence.
}

\item{
General relativistic effects such as time delay, lensing, and precession could
become important and affect optical/UV continuum and reverberated IR emission
differently.
}

\item{ 
Our predictions for reverberated periodic continuum hold true for any source
with a periodic component. Addition of any non-periodic component such as
noise, does not changes our findings, but rather adds an additional IR
signature on top of the one presented here. We expect that quasi-periodic
signatures will create reverberated IR light curves that can be treated as a
perturbation from those of the periodic case.
}

\end{itemize}

\subsection{Conclusion}

We have developed models to compute IR light curves that result from the
reverberation of a periodic optical/UV source by a surrounding dusty torus.
For the first time, we have considered the reverberation of periodic
optical/UV sources, with either isotropic luminosity variations, or
anisotropic source variability induced by the relativistic Doppler boost. The
latter setup resembles a rotating, forward-beamed light-house surrounded by
fog. We found a number of differences from the case of reverberation of an
isotropically variable source, and applied our findings to the IR emission
from the MBHB candidate PG~1302-102.  We found that dust tori are constrained
to be thin, and that the present data already places non-trivial constraints
on the inclination angle of the torus in the Doppler case, requiring the torus
to avoid being nearly face on.

The type of models described here can be applied to
fit the IR and optical/UV light curves of the growing list of MBHB
candidates \citep{Graham+2015b, Charisi+2016, Jun:2015}, aiding in
their interpretation, and could also aid in identifying new MBHB
candidates.

\section*{Acknowledgments} 

We thank Hyunsung Jun for providing the WISE data from his analysis, and
Hyunsung Jun, Daniel Stern, Arlin Crotts, Jules Halpern, Andrew MacFadyen,
Jeffrey J. Andrews, Adrian Price-Whelan, and James Guillochon for useful
discussions. DJD thanks Janna Levin and Pioneer Works for providing a
stimulating work space during the completion of this work. The authors
additionally thank the anonymous referee for his/her comments. This publication
makes use of data products from WISE, a joint project of the University of
California, Los Angeles, and the Jet Propulsion Laboratory/California
Institute of Technology, funded by NASA. This publication makes use of data
products from NEOWISE, which is a project of the Jet Propulsion
Laboratory/California Institute of Technology. NEOWISE is funded by the
National Aeronautics and Space Administration. Financial support was provided
from NASA through Einstein Postdoctoral Fellowship award number PF6-170151
(DJD) and NASA ATP grants NNX11AE05G and NNX15AB19G (ZH). ZH also gratefully
acknowledges support from a Simons Fellowship for Theoretical Physics.

\bibliographystyle{mnras}
\bibliography{refs}

\appendix

\section{Analytic Approximation to Doppler relative IR amplitudes} 
\label{A:DopApprox}

Throughout the text we approximate the Doppler dust echo using a
solution where the illuminating source has an effective spectral slope
of $\alpha_{\tot}=4$.  This is because this allows a simple, exact
analytic formula for the light curve and hence the relative amplitude
and phase, which we have shown agrees well with numerical solutions
for the more relevant ``bolometric'' Doppler spectral slope of
$\alpha_{\nu}\approx -1$  (see Fig.~\ref{Fig:Dop_alph}).

Although not used for the calculations in this paper, here we give
approximate analytic solutions for the relative amplitudes in the
$\alpha_{\tot}=-1$ case, which may be useful in future studies.

For small binary orbital velocity and when the binary is face on, the
approximate solutions presented here approach the $\alpha_{\tot}=4$
solutions used in the main text.

For a face on torus ($J=\pi/2$), we wish to evaluate the integral
\begin{equation}
\int^{2 \pi}_{0}\int^{\pi-\theta_T}_{\theta_T} \frac{\sin{\theta}}{\left(1 - v_{||}(t, t_{\em}(\theta,\phi),\theta,\phi)/c \right)^{4} } d\theta d \phi
\end{equation}
where $v_{||}$ is a function of $\phi$ and $\theta$ given by Eq.
(\ref{Eq:vlos}). Expanding the integrand in a binomial series to second order
in $\beta$ (the orbital velocity in units of the speed of light $c$),
and choosing a convenient observer position so that $t_{\em} = t -
t_d\left(1 - \cos{\theta} \right)$, we can solve analytically for the time
dependent light curve. Because we already have the optical/UV light
curve, we can find the maximum and minimum of both the IR and the UV light curves to find the amplitude of each.
Taking the ratio of amplitudes, we then find
\begin{eqnarray}
&&\frac{A_{\rm{IR}}}{A} = \nonumber \\
%
&&6 \gamma^4 
    \left( a -1 \right)^4
    \left[
    2 x \left( \cos{\theta_T} \left( 15 a \cos{\left[2 x \cos{\theta_T}\right]}
    +8 x \cos{\left[x \cos{\theta_T}\right]} \right)
     \right. \right. \nonumber \\
    &&- \left. \left. 8 \sin{\left[x \cos{\theta_T}\right]} \right)- 5 a \left( 3 x^2 \cos{2 \theta_T} + x^2 - 3 \right)
    \sin{\left[2 x \cos{\theta_T}\right]} 
    \right] / \nonumber \\
%
    && \left\{ 
    x \left( a - 2 \right) 
    \left[ a \left( a - 2 \right)  + 2 \right] \right. \nonumber \\
    && \left. \left[
    x \cos{\theta_T} \left(-24 -85 \beta^2 \right) x \beta 
        \left( 
        5 \beta x \left( 6 \cos{2I} \sin^2{\theta_T} + \cos{\theta_T} \right) \right.    \right.  \right. \nonumber \\
        &&+ \left. \left. \left. 48 \cos{I} \cos{\left[x \cos{\theta_T}\right]}
        \right)
        - 48 a \sin{\left[x \cos{\theta_T}\right]}
    \right]
    \right\} \\
    &&a \equiv \beta \cos{I} \qquad x \equiv \Omega t_d  \quad (\mbox{Face-on} \ \mbox{Torus}).\nonumber   
\end{eqnarray}
While $A=a$ for the $\alpha_{\tot}=4$ case and for $\gamma\rightarrow 1$, this is not true in general.
For a binary which is also face on ($I=\pi/2$) this solution simplifies considerably to
\begin{equation}
\frac{A_{\rm{IR}}}{A} = \frac{3}{3+10 \beta^2} \left[ \frac{ \rm{sinc}{ \left( \Omega t_d \cos{\theta_T} \right)  }    }{\Omega t_d} - \frac{ \cos{ \left(\Omega t_d \cos{\theta_T} \right) } }{ \Omega t_d } \right],
\end{equation}
which is the $\alpha_{\tot}=4$ exact solution presented in the main text to
first order in $\beta$. Figure \ref{Fig:Dop_alph} of the main text compares the
$\alpha_{\tot}=-1$, $\alpha_{\tot}=4$, and numerical solutions.

\begin{figure*}
\begin{center}$
\begin{array}{c c c}
\includegraphics[scale=0.24]{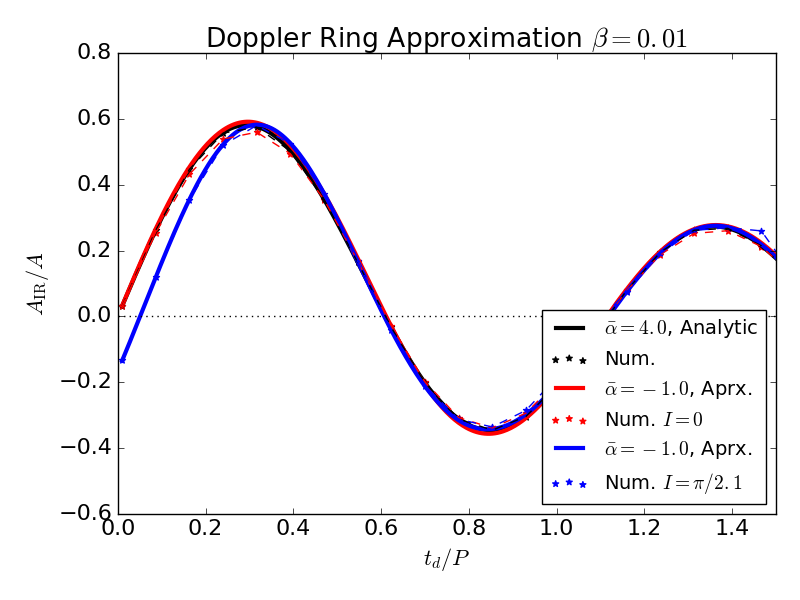} & 
\includegraphics[scale=0.24]{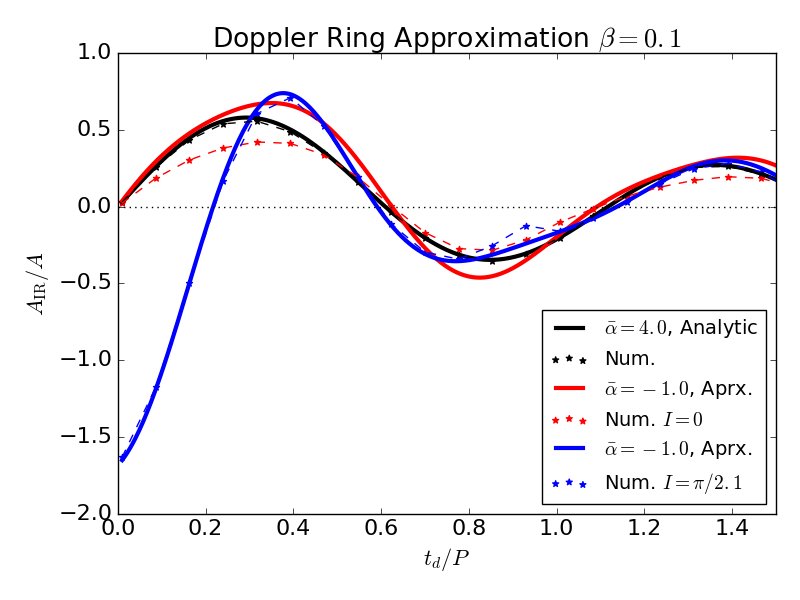}
\end{array}$
\end{center}
\caption{Comparison of the approximate solutions to the edge on Doppler ring when $\alpha_{\tot}=-1$ compared to numerical solutions and the exact analytic solution for $\alpha_{\tot}=4$.}
\label{Fig:DopRingApprox}
\end{figure*}

We also find the analog of Eq. (\ref{Eq:ISO_Ring}) for an edge-on dust ring in the Doppler case, but with the broad-band ``bolometric'' $\alpha_{\tot}=-1$. Following the same procedure as above, expanding to second order in $\beta$, we find
\begin{eqnarray}
\frac{A_{\rm{IR}}}{A} &=& \gamma^4 2 \sec{I} (a+1)^4 \left[4 x \cos{I} 
\left(
    5 a J_0(2x) + 2 J_1(x)
\right) \right. \nonumber \\
&-&\left. 5 \beta J_1(2x) (\cos{2I} + 3) \right] / \nonumber \\
&& \left\{ x \left[ a^3 + 4 a^2 + 6a + 4\right] \right. \nonumber \\
 && \left. \left[ 5\beta^2 \cos{2I} +15 \beta^2 -8a J_1(x) +4\right] \right\} \\
&&a \equiv \beta \cos{I} \qquad x \equiv \Omega t_d  \quad (\mbox{Edge-on} \ \mbox{Ring}),\nonumber
\end{eqnarray}
where $J_0(z)$ and $J_1(z)$ are the zeroth- and first-order Bessel functions of the first kind. Figure \ref{Fig:DopRingApprox} plots the $I=0$ and $I=\pi/2$ limits of this approximation compared with the numerical evaluation and the $\alpha_{\tot}=4$ analytic solution, for different values of $\beta$.

We find that 
\begin{enumerate}
\item The $\alpha_{\tot}=-1$, $I=0$ analytic approximation does not match well the numerical solution, while the $\alpha_{\tot}=-1$, $I=\pi/2$ analytic solution does.
\item Still, the $\alpha_{\tot}=-1$ solutions do not differ greatly from the $\alpha_{\tot}=4$ solution unless $I\rightarrow \pi/2$. 
\item The $\alpha_{\tot}=-1$, $I\rightarrow\pi/2$ solutions show new behavior. As $I\rightarrow\pi/2$, and for increasing $\beta$, the $\alpha_{\tot}=-1$ relative amplitudes blow up to infinity starting at $t_d/P =0$ and propagating outwards in $t_d/P$. This may be because the optical/UV amplitude is approaching zero as the binary approaches face on.
\end{enumerate}
Based on these results, and because we are not concerned with near face-on binaries in the case of PG~1302-102, we use the exact $\alpha_{\tot}=4$ solutions presented in the main text for vetting PG~1302-102 models.

\section{MCMC fits}
\label{A:MCMC}

In each case below we use the \textsc{emcee} code \citep{DFM:2013} to find
best-fit model parameters to the data. In the main text and in
Tables \ref{Table:SinParams} and \ref{Table:IRmeasures}, we quote our results
in terms of a maximum likelihood value $\pm$ the values which exclude the
outer $15\%$ of the likelihood.

\subsection{Sinusoid fit}
\label{A:SinFit}
For each of the V-band, W1, and W2 time series we first fit a sinusoid by minimizing
\begin{equation}
\chi^2\equiv\sum_{i} \frac{\left[ \rm{mag}(t_i) - \mathcal{A} \sin{\Omega \left( t_i - t_0 \right) } + \mathcal{B}\right]^2 }{ [\delta \rm{mag}(t_i)]^2}
\end{equation}

with respect to the three parameters $A$, $t_0$, and ${\rm   mag_0}$ in each
band, where $t_i$, mag$(t_i)$, and $\delta$mag$(t_i)$ are the observation
times, magnitudes, and errors. We impose the flat priors $A>0$ and $0 \leq t_0
\leq P$ and fix $\Omega$ to the best-fit value previously found by
\citet{Graham+2015a} corresponding to a rest-frame period of $P=1473.72$ days
(or observed $P_{\rm obs}=1884$ days) from $V$ band data.

\begin{figure*}
\begin{center}$
\begin{array}{c c c}
\includegraphics[scale=0.21]{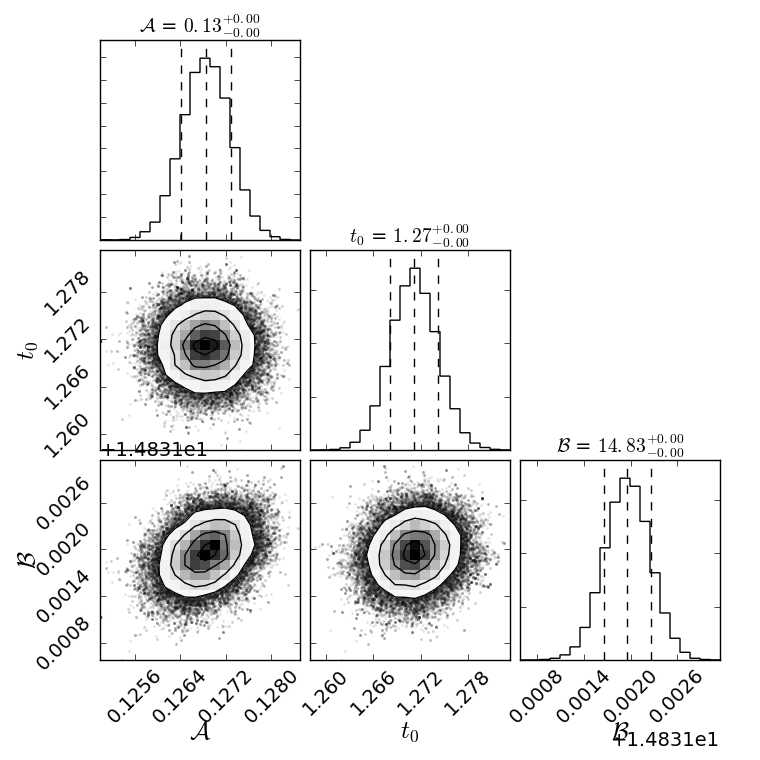} & 
\includegraphics[scale=0.21]{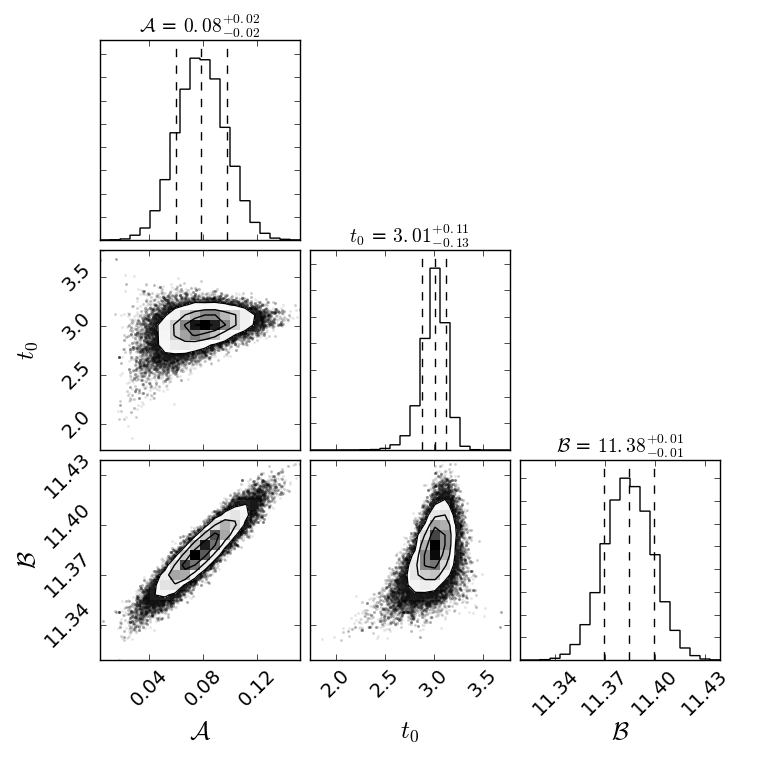} &
\includegraphics[scale=0.21]{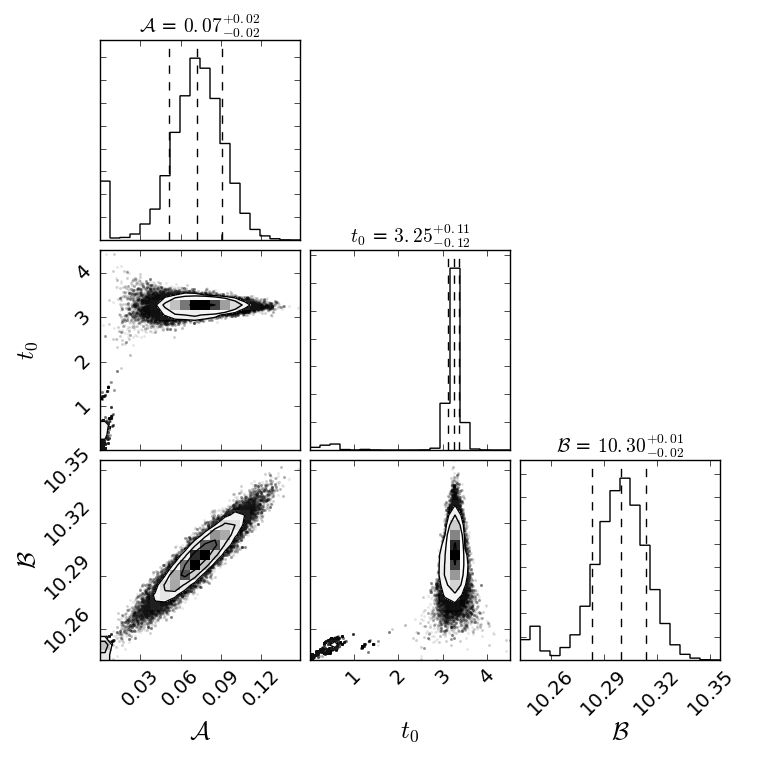}
\end{array}$
\end{center}
\caption{
MCMC sampling of posterior distributions for the three   sinusoidal fits
showing parameter correlations and posterior distributions for the V, W1,
and W2 bands, from top to bottom. We employ 48 walkers on 4096 step chains.
The phases are written in multiples of a cycle rather than in
rest-frame days, and all relative to zero. The 15, 0.5, and 0.85 quartiles are
drawn as dotted lines and quoted above each column.
}
\label{Fig:App:SinFit}
\end{figure*}

Figure \ref{Fig:SinFit} shows the best-fit sinusoids with this fixed
period. The black data points and error bars are the epoch-averaged
magnitudes and standard deviations used in the fit, while the coloured
points are the unbinned data. Figure \ref{Fig:App:SinFit} shows the
sampling of the posterior distribution.

\begin{figure*}
\begin{center}$
\begin{array}{c c c}
\hspace{-20pt}
\includegraphics[scale=0.16]{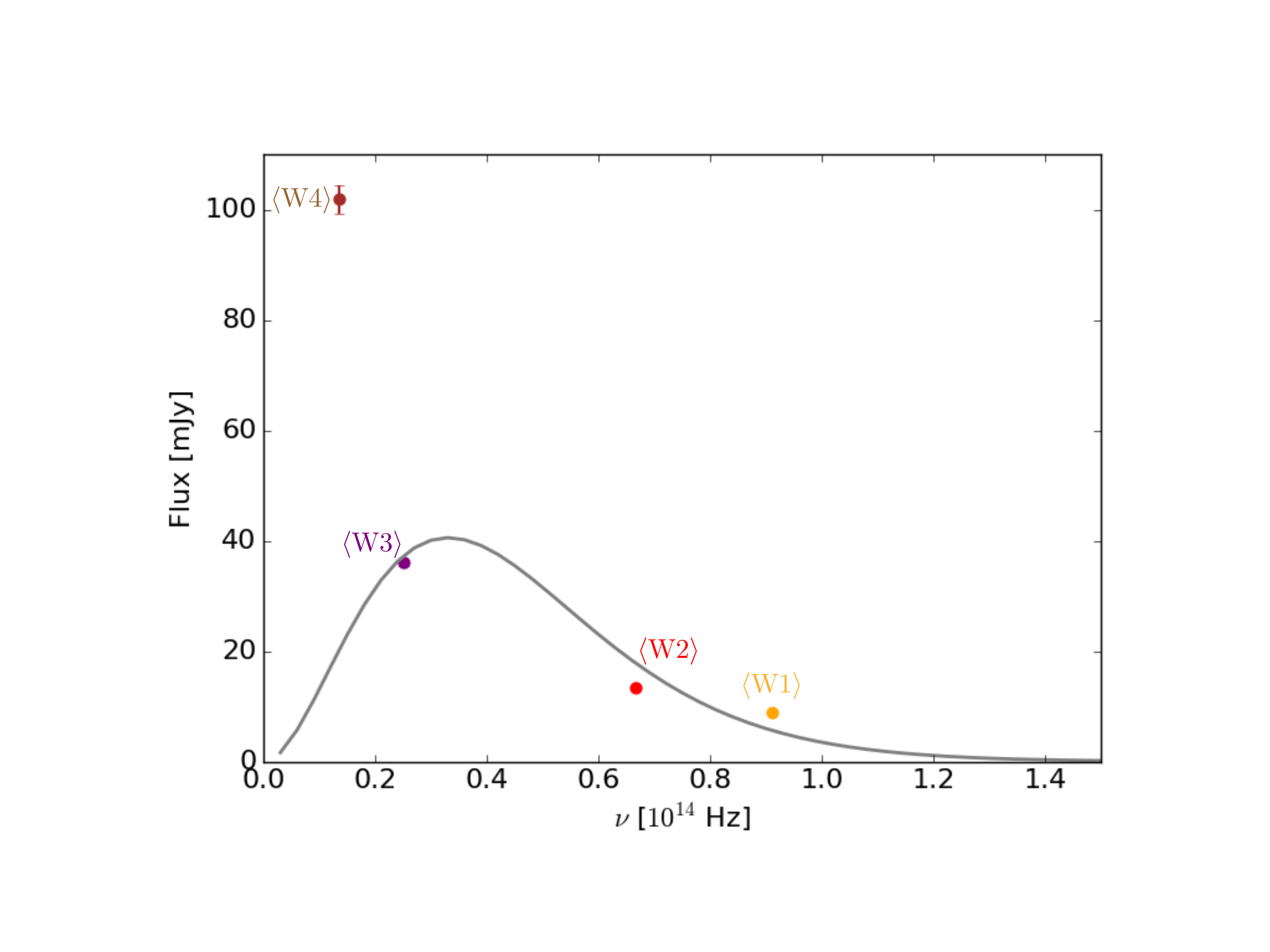}  &
\includegraphics[scale=0.16]{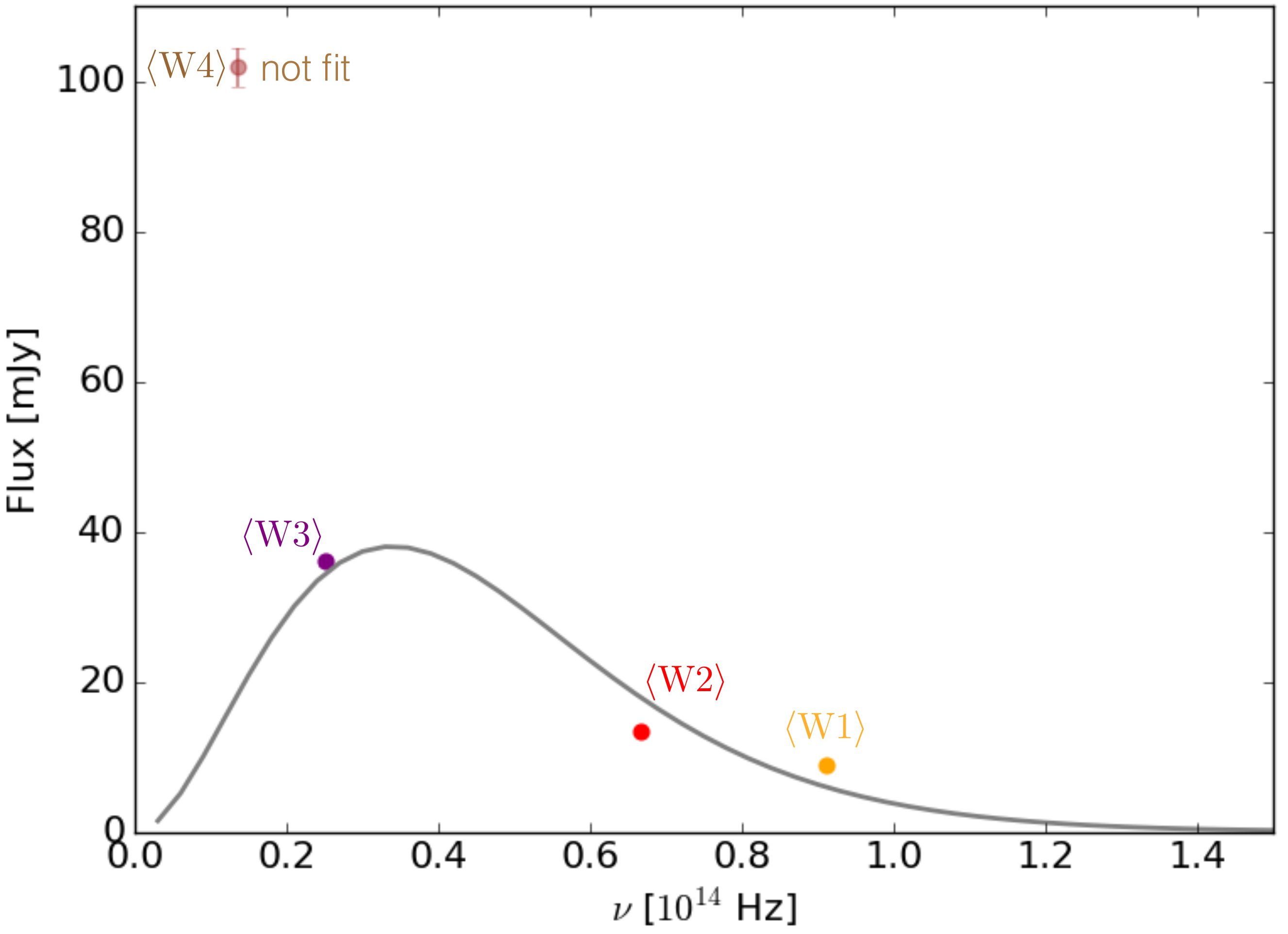}  &
\includegraphics[scale=0.16]{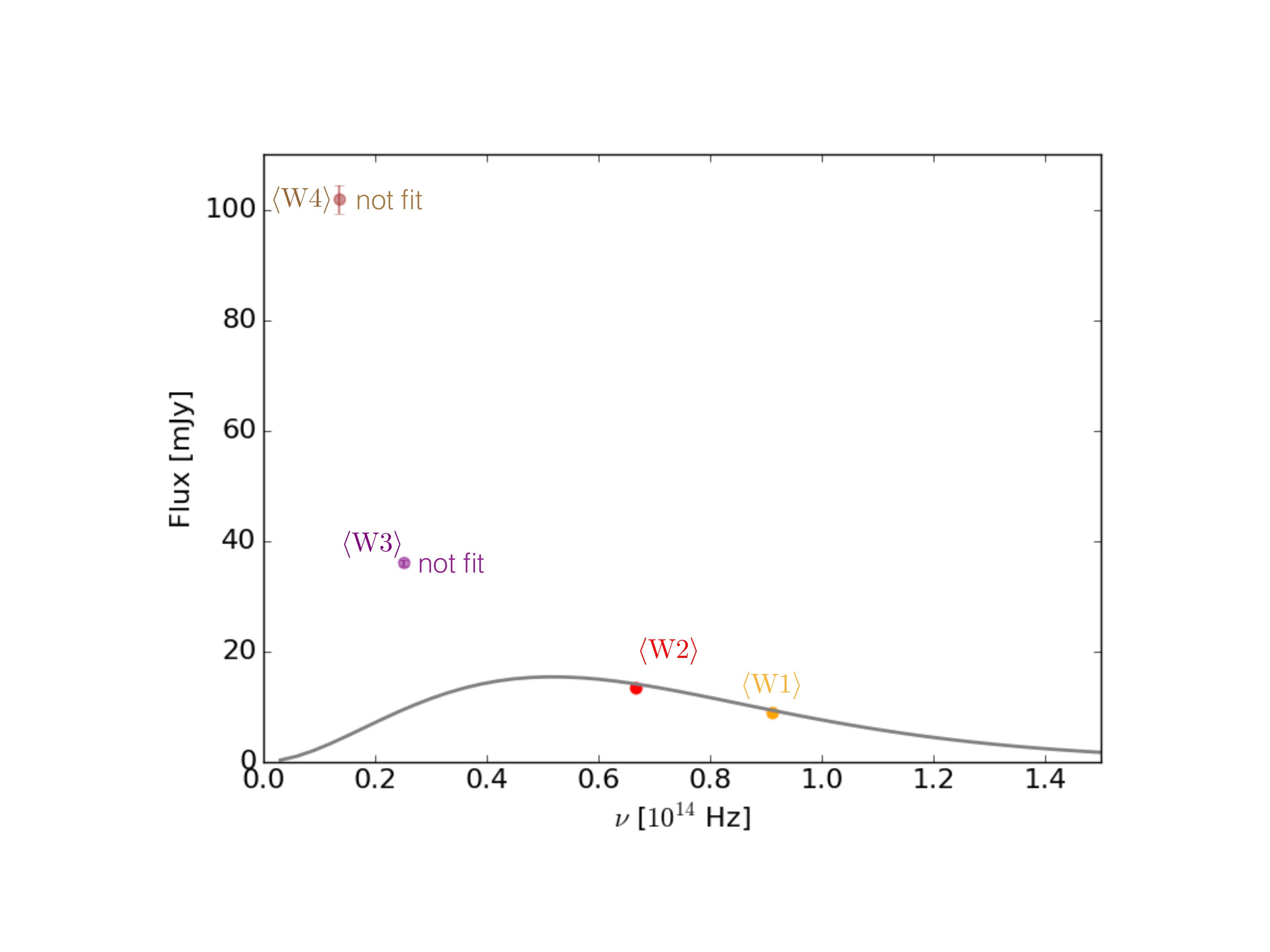}  \\
\end{array}$
\end{center}
\caption{
The best-fit blackbody spectra to the average WISE fluxes with one standard
deviation errors (error bars are smaller than data points for the W1, W2, and
W3 bands). From left to right, all four WISE bands are fit, only W1 through W3
are fit, and only W1 and W2 are fit.
}
\label{Fig:BBFit}
\end{figure*}

\subsection{Blackbody SED fit} 
\label{A:BBfit} 
The best-fit blackbody spectral
energy distribution is found for three different cases: using the time-
averaged fluxes from all four WISE bands, only the fluxes from the W1, W2,
and W3 bands, and also using only the W1 and W2 bands. The average WISE fluxes
are, $F_{\rm{W1}} = 8.83 \pm 0.18$ mJy, $F_{\rm{W2}} = 13.33 \pm 0.26$ mJy,
$F_{\rm{W3}} = 36.07 \pm 0.56$ mJy, and $F_{\rm{W4}} = 101.859 \pm 2.62$ mJy
(1$\sigma$ errors). We minimize 
\begin{equation}
\chi^2\equiv\sum_{\nu=W1,W2,...} \frac{\left(F_{\nu} -   X \frac{ B_{\nu}[T_d]}{T^4_d}\right)^2 }{(\delta F_{\nu})^2}
\end{equation}
with respect to two parameters: the dust temperature $T_d$ and $X$, a combination of the dust covering fraction $\cos{\theta_T}$ and total source luminosity after the bolometric correction,
\begin{equation}
X \equiv 4 \pi \cos{\theta_T} \left[ \frac{\rm{BC} \ L_V}{16 \pi \sigma_{\rm{SB}} d^2 } \right],
\end{equation}
where $d$ is the luminosity distance to the source and $L_V$ is the V-band luminosity.

The above model flux derives from the assumption that the grains are
optically thin to IR. Then the flux from each grain is $\pi B_{\nu}[T_d]
(a_{\rm{eff}}/d)^2$ and there are $N = \Sigma_d
\int^{2\pi}_0\int^{\pi-\theta_T}_{\theta_T} R^2_d \sin{\theta} \ d\phi \
d\theta = 4 \pi R^2_d \cos{\theta_T}$ grains. For an optically thick grain
distribution $\Sigma \rightarrow (\pi a^2_{\rm{eff}})^{-1}$ and we obtain
\begin{equation}
F^{\rm{model}}_{\nu} = 4 \pi \cos{\theta_T} (R_d/d)^2 B_{\nu}[T_d] 
\end{equation}
which equals $ X  B_{\nu}[T_d]/T^4_d$ upon imposing the thermal equilibrium constraint
\begin{equation}
R_d = \left[ \frac{\rm{BC} \ L_V}{16 \pi \sigma_{\rm{SB}} T^4_d} \right]^{1/2}.
\end{equation}

Once we fit for $T_d$ and $X$, we impose an empirically motivated range of the
bolometric correction \citep{RichardsQBCs:2006} which limits the possible
range of the covering fraction that can reproduce the total observed IR
luminosity. We choose BC$=9\pm3$ to obtain the range of $\cos{\theta_T}$
presented in Table \ref{Table:IRmeasures}.

\label{lastpage}
\end{document}